\documentclass[a4paper,10pt,twocolumn,amsmath,amssymb]{revtex4}
\usepackage[utf8]{inputenc}
\usepackage{graphicx}
\usepackage{bm}

\begin{document}

\preprint{AIP/123-QED}

\title{Phase and frequency linear response theory for hyperbolic chaotic oscillators}
\author{Ralf T\"onjes}
\affiliation{Institute of Physics and Astronomy, Potsdam University, 14476 Potsdam-Golm, Germany
}%
\altaffiliation{toenjes@uni-potsdam.de}
\author{Hiroshi Kori}%
\affiliation{Department of Complexity Sciences and Engineering,
Univeristy of Tokyo, Kashiwa, 277-8561 Chiba, Japan
}%
\altaffiliation{kori@k.u-tokyo.ac.jp}
\date{\today}

\begin{abstract}
We formulate a linear phase and frequency response theory for hyperbolic flows, which generalizes phase response theory for autonomous limit cycle oscillators to hyperbolic chaotic dynamics. The theory is based on a shadowing conjecture, stating the existence of a perturbed trajectory shadowing every unperturbed trajectory on the system attractor for any small enough perturbation of arbitrary duration and a corresponding unique time isomorphism, which we identify as phase, such that phase shifts between the unperturbed trajectory and its perturbed shadow are well defined. The phase sensitivity function is the solution of an adjoint linear equation and can be used to estimate the average change of phase velocity to small time dependent or independent perturbations. These changes of frequency are experimentally accessible giving a convenient way to define and measure phase response curves for chaotic oscillators. The shadowing trajectory and the phase can be constructed explicitly in the tangent space of an unperturbed trajectory using co-variant Lyapunov vectors. It can also be used to identify the limits of the regime of linear response. 
\end{abstract}

\maketitle

{\bf
Phase response curves are a powerful tool to predict and analyze synchronization of weakly forced or coupled oscillators. The state of chaotic oscillators, however, is not characterized by a unique geometric phase. Even if a geometric phase is imposed, the phase difference between two identical chaotic oscillators is not asymptotically constant or even bounded, whereas phase response is commonly measured as the asymptotic phase shift caused by a single pulsed perturbation. In this report, we reinterpret phase as a time isomorphism rather than a geometric angle. This allows us to generalize linear phase response theory to chaotic oscillators, as well as to predict and measure the phase response via experimentally accessible frequency shifts.
}
\section{Introduction}
Synchronization, the adaptation of frequencies of self-sustained oscillators to a driving force, plays a vital role in many systems, ranging from biological and chemical systems to artificial devices \cite{pikovsky2003synchronization, winfree2001geometry, kuramoto2003chemical, glass2001synchronization}, and its understanding is essential for prediction and control of collective behavior. Synchronization can manifest in many forms, 
weakly as a resonance in periodically forced stochastic oscillators \cite{tonjes2011synchronization}, or more strongly as a locking of
oscillation frequencies, phases, complete or generalized synchronization
\cite{rulkov1995generalized}.
Phase synchronization in weakly coupled or weakly forced, autonomous limit cycle oscillators can be understood by linear phase response theory, which describes the evolution of a phase $\varphi=\varphi(t)$, defined on a circle with the perimeter of its natural period $T_0 = 2\pi/\omega_0$, in linear order of a perturbation $\varepsilon\vec{p}(\varphi,t)$ as
\begin{equation}\label{Eq:LCPhaseResponse}
	\dot\varphi = 1 + \varepsilon \vec{Z}(\varphi)\cdot\vec{p}(\varphi,t).
\end{equation}
Note that in this convention, phase $\varphi$ has the dimension of time. 
While Eq.\,\eqref{Eq:LCPhaseResponse} describes the change of phase
velocity in linear order of $\varepsilon$, the equation is nonlinear in
$\varphi$ and even small perturbations can aggregate to nonlinear
synchronization effects.
Equations like \eqref{Eq:LCPhaseResponse} are sometimes referred to as
Winfree type phase equations in recognition of his unifying works in
mathematical biology \cite{winfree1967biological,winfree2001geometry}. The function
$\vec{Z}(\varphi)$ is called phase sensitivity function and its
components are proportional to phase response curves (PRCs). The PRCs
essentially determine a system's synchronization behavior and are used
in mathematical modeling of weakly coupled oscillators across
scientific disciplines from biology, in particular neuroscience
\cite{hoppensteadt2012weakly} and chronobiology \cite{winfree2001geometry}, chemistry, ecology
to electrical engineering and many others \cite{pikovsky2003synchronization}. Based on
the PRCs it is possible to design perturbation protocols that can
stabilize or destabilize various collective modes in ensembles of
oscillators including complete synchronization, clustering and the asynchronous
state \cite{kiss2007engineering} or perform other control tasks in an optimal way
\cite{zlotnik2013optimal}. 
In this paper we will discuss if and in what sense Eq.\,\eqref{Eq:LCPhaseResponse} can be used for more general dynamics than limit cycle oscillators. The key is to note that phase in Eq.\,\eqref{Eq:LCPhaseResponse} has the dimension of time and evolves as time in an unperturbed system \cite{freitas2018phase}. Thus, instead of interpreting phase as a geometric angle-like variable we can reinterpret phase as a time isomorphism $\varphi=\varphi(t)\in\mathbb{R}$ defined by Eq.\,\eqref{Eq:LCPhaseResponse} which parameterizes a typical trajectory $\vec{x}_0(\varphi)$ on a hyperbolic attractor. Indeed, in the following we will adopt the viewpoint that phase is time in the unperturbed system, i.e. 
\begin{equation}\label{Eq:x0_Dyn}
    \frac{d\vec{x}_0}{d\varphi} = \vec{f}(\vec{x}_0).
\end{equation}
For stable limit cycle oscillators the distance between a perturbed trajectory $\vec{x}(t)$ with
\begin{equation}\label{Eq:xpert_dyn}
    \dot{\vec{x}} = \vec{f}(\vec{x}) + \varepsilon \vec{p}(\vec{x},t)
\end{equation}
and the phase shifted unperturbed trajectory $\vec{x}_0(\varphi(t))$
is bounded by $O(\varepsilon)$ for all times and arbitrary perturbations. Then Eq.\,\eqref{Eq:LCPhaseResponse} with $\vec{Z}(\varphi)=\vec{Z}(\vec{x}_0(\varphi))$ and $\vec{p}(\varphi,t)=\vec{p}(\vec{x}_0(\varphi),t)$ predicts the phase velocity in linear order of $\varepsilon$.
Throughout the paper we assume $\vec{x}_0=\vec{x}_0(\varphi)$ to be a solution of the unperturbed system \eqref{Eq:x0_Dyn} evolving on an invariant set, e.g. a limit cycle or a chaotic attractor. Vector fields, such as the phase sensitivity can be expressed as functions of space or of time $\vec{Z}=\vec{Z}(\vec{x}_0)=\vec{Z}(\varphi)$ with respect to the points of the trajectory. Note, that the scalar $\varepsilon$ in \eqref{Eq:xpert_dyn} quantifies to the linear order the strength of any perturbation. Such a perturbation does not need to be additive but can be applied to a system parameter, as well. E.g. with $\vec{f}=\vec{f}(\vec{x},\mu)$ and $\mu=\mu_0+\varepsilon\Delta\mu$ \eqref{Eq:xpert_dyn} takes the form $\dot{\vec{x}} = \vec{f}(\vec{x},\mu_0) + \varepsilon\Delta\mu\partial_\mu\vec{f}(\vec{x},\mu_0)$.
\\ \\
In Sec.\,\ref{Sec:Review} we review the classic experimental and numerical methods to obtain the phase sensitivity for autonomous limit cycle oscillations. In Sec.\,\ref{Sec:Results} we generalize these methods to hyperbolic chaotic oscillators. We show in Sec.\ref{sec:freqresp} how our re-interpretation of phase as time in the unperturbed system can be used to define phase sensitivity from the frequency response of an oscillatory system. In Sec.\,\ref{sec:geometric} we improve on a well established linear least squares method \cite{schwabedal2012optimal} to define approximate Isochrons for chaotic oscillators. The main contribution of this paper in Sec.\,\ref{sec:Lyavecs} is the proposal to use covariant Lyapunov vectors \cite{ginelli2007characterizing} to define the phase sensitivity function for hyperbolic chaotic oscillators. We test this proposal in numerical examples in Sec.\,\ref{Sec:Examples}.
\section{Phase response functions of limit cycle oscillators}\label{Sec:Review}
There are three common and equivalent
approaches to obtain the linear PRCs of autonomous limit cycle oscillators as
described in the works of Winfree, Kuramoto and Malkin. These are based
respectively (i) on asymptotic phase or time shifts caused by single impulses at a
prescribed phase \cite{winfree1967biological}, (ii) on calculating the gradient of isochrons which are parameterized by the periodic phase \cite{kuramoto2003chemical}, and (iii) on the solution to an adjoint linearized equation
\cite{Malkin1949Methods, izhikevich2007dynamical}. The direction and amplitude of $\vec{Z}(\vec{x}_0)$ at a point $\vec{x}_0$ of a limit cycle follow from two geometric considerations :
$\vec{Z}$ must be perpendicular to the stable invariant manifold since perturbations on this manifold do not lead to phase shifts. Secondly, a perturbation in the direction of the flow advances the phase by an amount inversely proportional to the flow velocity, i.e. $\vec{Z}(\vec{x})\cdot\vec{f}(\vec{x}) = 1$.
\subsection{Measuring time shifts}
The first method is an experimental approach and requires no mathematical model of the system dynamics. Deviations from a stable limit cycle caused by a small, single pulsed perturbation $\vec{p}=\Delta\vec{x}\delta(t-t_0)$ at a phase $\varphi_0=\varphi(t_0^-)$ decay exponentially fast. The instantaneous phase shift $\varphi(t_0^+)-\varphi(t_0^-) = \Delta \varphi = \varepsilon \vec{Z}(\varphi_0)\cdot\Delta\vec{x}$ according to Eq.\,\eqref{Eq:LCPhaseResponse} remains constant afterwards and can be measured as a permanent time shift between the perturbed and an unperturbed system signal. $Z_\Delta(\varphi_0) = \lim_{\varepsilon\to 0}\Delta \varphi/\varepsilon$ is a phase response function. The index $\Delta$ stands for any experimentally realizable pulsed perturbation, either in the dynamic variables or in the system parameters. E.g., kicking the system in a single component of the state variable $\vec{x}$, i.e. replacing $x$ by $x+\varepsilon\Delta x$, will result in a time shift $\Delta\varphi/\varepsilon \to Z_x \Delta x = Z_{\Delta}$. In control problems the system state may not be directly accessible and a system parameter $\mu$ may only vary within practical limits. In this case it is impossible to apply a delta kick and a localized parametric forcing $\mu=\mu_0+\varepsilon\Delta\mu(\varphi)$ over a finite time interval $[\varphi_0-\tau,\varphi_0+\tau]$ and finite strength must be applied which results in a time shift
\begin{equation}
    \frac{\Delta\varphi}{\varepsilon}\to Z_\Delta(\varphi_0) = \int_{\varphi_0-\tau}^{\varphi_0+\tau}\vec{Z}(\vec{x}_0(\varphi))\cdot\partial_\mu\vec{f}\cdot\Delta\mu(\varphi)\,d\varphi.
\end{equation}
\subsection{Isochrons}
Isochrons (or isophases) $I_\varphi$ are invariant manifolds under the system propagation over one oscillation period $T_0$. They intersect the limit cycle in one point $\vec{x}_0(\varphi)$ which is an attracting fixed point of the time $T_0$ forward map on $I_\varphi$. All points on an isochron have the same phase $\varphi(I_\varphi)=\varphi(\vec{x}_0)$ and the same phase velocity $\dot\varphi(I_\varphi)=\dot\varphi(\vec{x}_0)=1$. Thus phase is defined everywhere in the basin of attraction of the limit cycle as a scalar field $\varphi=\varphi(\vec{x})$. Phase response is not restricted to small perturbations of a system close to the limit cycle \cite{kuramoto2003chemical,wilson2018greater}. The phase sensitivity is given as the gradient $\vec{Z}(\vec{x})=\vec{\nabla}\varphi(\vec{x})$, which is orthogonal to the isochrons and with $\dot\varphi = 1$ follows $\dot\varphi = \vec{\nabla}\varphi\cdot\dot{\vec{x}}=\vec{Z}\cdot\vec{f}=1$ everywhere. 
\subsection{Malkin's adjoint method}
Malkin's method considers deviations from the limit cycle only to the linear order. Here isochrons are linear subspaces in the tangent space at each point $\vec{x}_0$ of the limit cycle. Vectors $\vec{h}$ in the tangent space evolve under the periodic action of the system Jacobian matrix $\textrm{J}_f(\varphi)=\textrm{J}_f(\vec{x}_0(\varphi))$ with $(\textrm{J}_f)_{ij} = \partial f_i/\partial x_j$ along the limit cycle as $d\vec{h}/d\varphi=\textrm{J}_f\vec{h}$ . Invariance under system propagation over one period means that an invariant subspace is spanned by Floquet vectors. Perturbations in the stable directions do not change the phase, whereas perturbations in the direction $\vec{f}$ of the flow do not decay. 
The co-vectorfield $\vec{Z}(\vec{x}_0)$ which is the unique solution of the adjoint linear equation
\begin{equation}\label{Eq:MalkinAdjoint}
    \frac{d}{d\varphi}{\vec{Z}} = - J_f^\top(\varphi) \vec{Z}  
\end{equation}
on the limit cycle normalized to $\vec{Z}\cdot\vec{f} = 1$ is orthogonal to the stable invariant subspace (see Sec.\ref{sec:shadow}) and therefore equal to the linear phase sensitivity \cite{Malkin1949Methods,izhikevich2007dynamical}. Malkin's adjoint method is the standard way to obtain the phase sensitivity numerically, when the linearization $\textrm{J}_f(\varphi)$ of the dynamics at the limit cycle $\vec{x}_0(\varphi)$ is available.

%
\section{Phase response for chaotic oscillators}\label{Sec:Results}
Since the discovery of chaotic phase synchronization
\cite{rosenblum1996phase} many heuristic approaches have been suggested to
generalize phase response theory to autonomous chaotic oscillators and
to define PRCs or phase coupling functions
\cite{pikovsky1997UPOs,josic2001phase,beck2003geometric,schwabedal2012optimal,rusin2011engineering,kurebayashi1theory}. The main
difficulty is, that due to mixing and chaotic phase diffusion, usually
no globally differentiable isochrons exist in chaotic oscillators. Phase
shifts caused by perturbations are not asymptotically constant and can
therefore not be measured in a unique way.
All three methods must and can be modified if one wants to apply them to chaotic oscillators. 
\\ \\
In the following we will distinguish time like phase $\varphi$ from an angle like geometric phase $\vartheta(\vec{x})$ which parameterizes a periodic foliation of the state space into Poincare sections $P_\vartheta=P_{\vartheta+2\pi}$ and is increasing monotonously ($d\vartheta/d\varphi>0$) along a trajectory $\vec{x}_0(\varphi)$.
In general a geometric phase $\vartheta_0$, e.g. reconstructed from a time series by Hilbert-transform or some other embedding technique, does not evolve uniformly. Such a geometric phase is called a protophase. For limit cycles a simple rescaling from an arbitrary protophase $\vartheta_0$ to a uniformly evolving geometric phase $\vartheta$ is always possible \cite{Kralemann2008}.
\subsection{Measuring frequency response}\label{sec:freqresp}
Eq.\,\eqref{Eq:LCPhaseResponse} describes a time isomorphism $\varphi=\varphi(t)$. Conversely, time as a function of phase evolves to the linear order in $\varepsilon$ as
\begin{equation}\label{Eq:TimeResponse}
    \frac{dt}{d\varphi}= \frac{1}{1+\varepsilon\vec{Z}\cdot\vec{p}} = 1-\varepsilon\vec{Z}(\varphi)\cdot\vec{p}(\varphi,t) + O(\varepsilon^2).
\end{equation}
For perturbations $\vec{p}=\vec{p}(\varphi)=\vec{p}(\vec{x}_0(\varphi))$ without explicit time dependence we can take the average over $\varphi$ corresponding to an average along an unperturbed trajectory $\vec{x}_0(\varphi)$ and obtain
\begin{equation}\label{Eq:Zp_projection}
    \frac{T_\varepsilon}{T_0} = 1 - \varepsilon \left\langle \vec{Z}(\varphi)\cdot\vec{p}(\varphi)\right\rangle_\varphi
\end{equation}
where $T_0=1/\nu_0$ and $T_\varepsilon=1/\nu_\varepsilon$ are the average periods, $\nu_0$ and $\nu_\varepsilon$ the frequencies of the unperturbed and of the perturbed system, respectively. 
Instead of measuring an asymptotic time shift caused by a single perturbation pulse it is also possible to measure the shift of the average oscillation period or frequency in linear response to a perturbation that only depends on the position on the attractor
\begin{equation}\label{Eq:FreqShift}
     \left\langle \vec{Z}(\varphi)\cdot\vec{p}(\varphi)\right\rangle_\varphi = \lim_{\varepsilon\to 0}\frac{1}{\varepsilon} \frac{T_0-T_\varepsilon}{T_0}
    = \lim_{\varepsilon\to 0}\frac{1}{\varepsilon} \left(\frac{\nu_\varepsilon}{\nu_0}-1\right).
\end{equation}
The phase sensitivity may be expanded into a set of vector fields $\vec{Z}=\sum_k z_k \vec{p}_k$ which are orthonormal under the scalar product on the left hand side such that
\begin{equation}
    z_k = \left\langle \vec{Z}(\varphi)\cdot\vec{p}_k(\varphi)\right\rangle_\varphi.
\end{equation}
Or the system is kicked with $\vec{p} = \Delta\vec{x} \sum_{i}\delta(t-t_i)$ everytime $t_i$ a Poincare section $P_\vartheta$ is crossed after one oscillation. Then from Eq.\,\eqref{Eq:LCPhaseResponse} follows that the average PRC on that Poincare section is
\begin{equation}\label{Eq:TimeShiftPRC}
    Z_\Delta(\vartheta) = \lim_{\varepsilon\to 0} \frac{1}{\varepsilon}(T_0-T_\varepsilon)
\end{equation}
if the limit exists, i.e. the chaotic system does have a linear response to the perturbation $\vec{p}$. All propositions for a phase sensitivity $\vec{Z}$ must be judged by comparing the predicted frequency shifts to measurements. The works \cite{rusin2011engineering,kurebayashi1theory} use frequency response to define such average or effective PRCs on Poincare sections $P_\vartheta$ constructed from the $T_0$ forward map.
\subsection{Optimizing a geometric phase}\label{sec:geometric}
A possible heuristic approach is to define isochrons as a family of Poincare sections $P_\vartheta$ parameterized by a geometric phase $\vartheta\in [0,2\pi)$ and optimize these surfaces under a set of constraints such that the variance of the return time is minimized \cite{schwabedal2012optimal}. Here, instead of the time domain, we perform the optimization in the frequency domain which has some advantages, as we will see. As in \cite{schwabedal2012optimal} we expand a geometric phase $\vartheta_\sigma(\vec{x})$ around a proto-phase $\vartheta_0(\vec{x})$ in the neighborhood of the attractor into an appropriate set of non-constant, differentiable functions $q_k(\vec{x})$
\begin{equation}\label{Eq:OptiPhaseAnsatz}
	\vartheta_{\sigma}({\vec x}) = \vartheta_0({\vec x}) + \sum_{k} \sigma_k q_k({\vec x}) \quad\textrm{mod}\quad 2\pi.
\end{equation}
e.g. Laguerre polynomials and spherical harmonics in spherical coordinates or Taylor polynomials and Fourier components in cylindrical coordinates. Given the vector fields $\vec{v}^{(l)}(\vec{x}_0)$ in the stable, unstable and $\vec{v}^{(0)}=\vec{f}$ neutrally stable directions on the attractor, we require the gradient $\vec{\nabla}\vartheta_\sigma$ to be orthogonal to the stable and unstable directions $\vec{\nabla}\vartheta\cdot\vec{v}^{(l\ne 0)}\approx 0$ and $\vec{\nabla}\vartheta\cdot\vec{f}=\dot\vartheta \approx \omega_0$. Indeed, such a vector field $\vec{Z}\parallel\vec{\nabla}{\vartheta}$ exists and is uniquely determined by the vectors $\vec{v}^{(l)}$. It can be used as phase sensitivity function in some sense, as we will discuss in the next section. As a finite sum of differentiable functions $\vartheta_0$ and $q_k$, the gradient $\vec{\nabla}\vartheta_\sigma$ of the geometric phase $\vartheta_\sigma$ is a differentiable approximation of $\vec{Z}$ such that $\vec{\nabla}\vartheta_\sigma\approx\omega_0\vec{Z}$ and
\begin{equation}\label{Eq:WLLS}
    \vec{\nabla}\vartheta_\sigma \cdot\vec{v}^{(l)} =
    \vec{\nabla}\vartheta_0 \cdot\vec{v}^{(l)} + 
    \sum_k \sigma_k \vec{\nabla}q_k \cdot\vec{v}^{(l)} = \omega_0\delta_{l0} + \eta_l.
\end{equation}
Applying the method of linear least squares to Eq.\,\eqref{Eq:WLLS} the coefficients $\sigma_k$ can be found which minimize the square norm of the deviations $\eta_l(\vec{x}_0)$ over all points on the attractor. Choosing $l=0$, i.e. $\vec{v}^{(0)}=\vec{f}$, we can include points, and calculate $\vec{f}$ there, which are close but not exactly on the attractor. Then $\vec{\nabla}\vartheta_\sigma\cdot\vec{f} = \dot\vartheta_\sigma\approx\omega_0$ will evolve approximately uniformly in the neighborhood of the attractor. The advantages over the method \cite{schwabedal2012optimal} of Schwabedal et al. are that the phase velocity of points which are not on the attractor can easily be calculated in contrast to the return times and that we can include additional linear constraints if the stable and unstable directions are available. 
Note that the lengths of the vectors $\vec{v}^{(l)}$ with $l\ne 0$ are arbitrary. Choosing them, e.g. in some relation to the flow velocity $\vec{f}$ makes Eq.\,\eqref{Eq:WLLS} a weighted linear least squares problem. Secondly, including all Lyapunov vectors in the linear least squares problem essentially results in a smooth geometric phase with a phase gradient that on the attractor approximates the theoretical phase sensitivity $\frac{1}{\omega_0}\vec{\nabla}\vartheta \approx \vec{Z}$.

\begin{figure}[!t]
\setlength{\unitlength}{1cm}
\begin{picture}(4.2,4.2)
\put(0,0){\includegraphics[height=4.0cm]{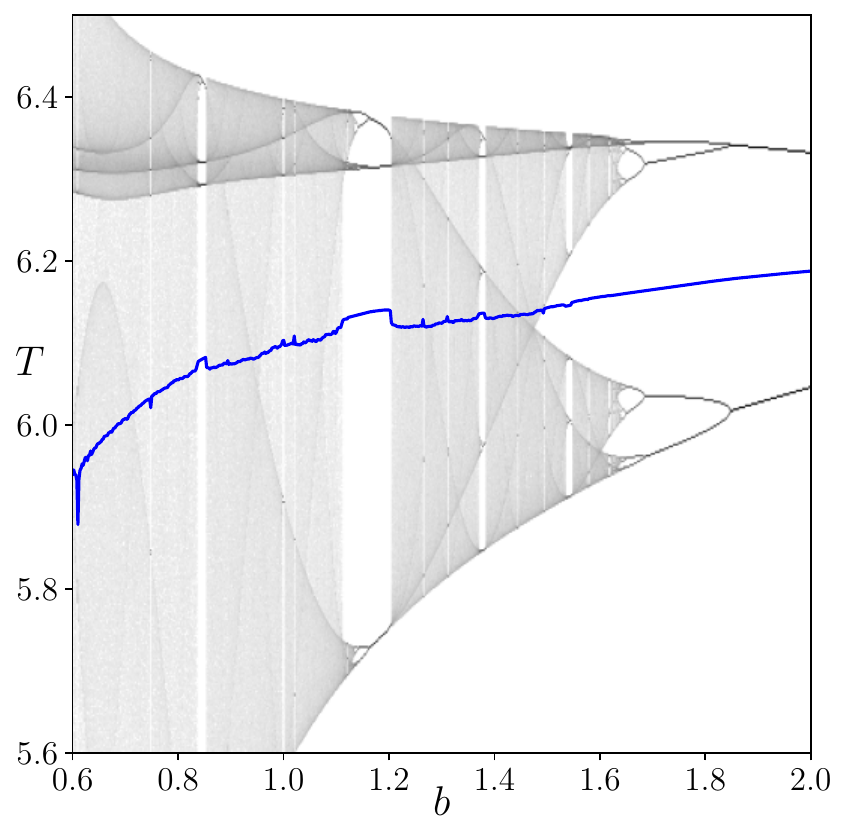}}
\put(-0.1,3.8){\bf (a)}
\end{picture}
\begin{picture}(4.2,4.2)
\put(-0.1,0){\includegraphics[height=4.2cm]{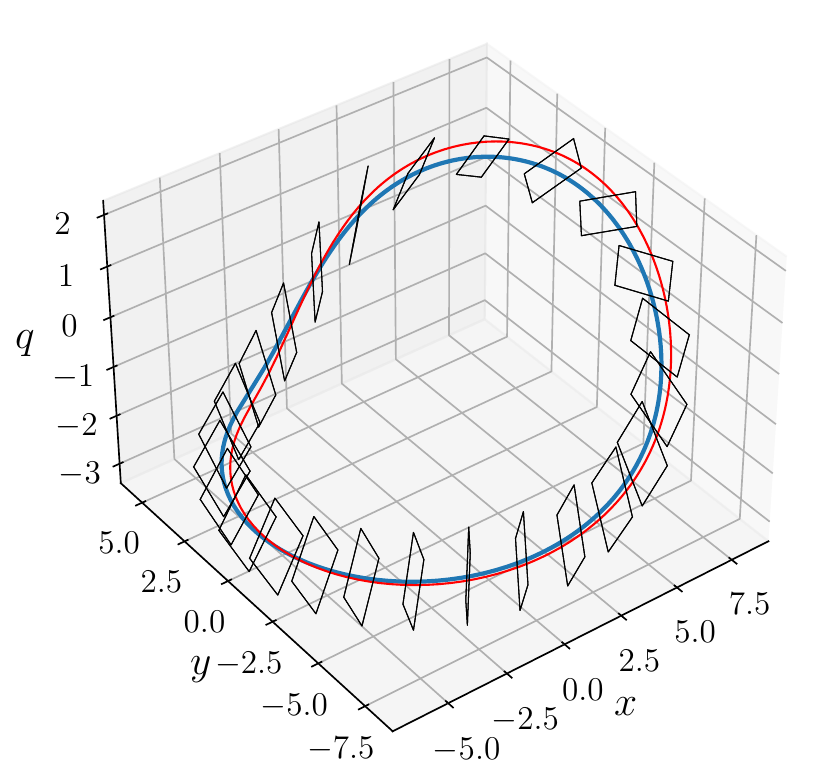}}
\put(0,3.8){\bf (b)}
\end{picture}
\caption{(a) The mean frequency of the chaotic Roessler oscillator (Eq.\,\eqref{Eq:RoesslerSystem}, $a=0.25, c=6.0$) is not a differentiable function of the system parameter $b$ at points of bifurcation. Shown are histograms of return times to the Poincare section $P_{\vartheta_0}$ at $\vartheta_0=\pi/3$ and the mean period (blue line) as functions of $b$. (b) Unstable periodic orbit (solid blue line) of the chaotic Roessler oscillator Eq.\,\eqref{Eq:RoesslerSystem} ($a=0.25, b=0.9, c=6.0$) with natural frequency $\omega_0=1.04$. The invariant linear subspaces under system propagation of one period (black polygons) are linear approximations of the UPO's isochrons. The red line is the linear approximation of the UPO's shadow under periodic forcing of $\varepsilon\sin(\Omega t)$ in the $x$-direction. The UPO was found via numerical root finding on a Poincare section, the stable and unstable directions by forward and backward integration, and the shadow was constructed with the method described in Sec.\ref{sec:shadow}. With $\varepsilon=0.4$ and $\Omega=1.07$ the shadowing trajectory is synchronized and phase locked to the forcing.}\label{Fig:UPOresp}
\end{figure}
\subsection{Using co-variant Lyapunov vectors}\label{sec:Lyavecs}
\begin{figure}[!t]
\setlength{\unitlength}{1cm}
\begin{picture}(4.2,4.2)
\put(-1,0){\includegraphics[height=4.0cm]{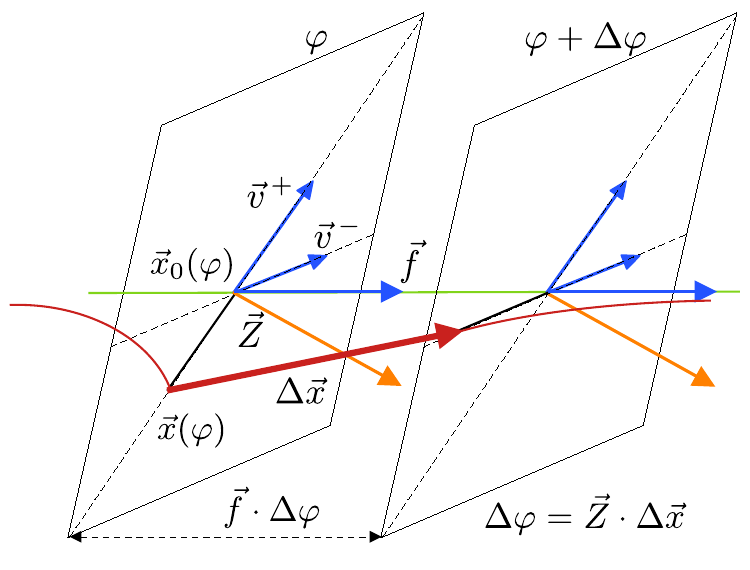}}
\end{picture}
\caption{Three dimensional schematics for the linear dynamics near a point $\vec{x}_0(\varphi)$ of an unperturbed trajectory (light green line). The subspace spanned by the stable and unstable directions (co-variant Lyapunov vectors $\vec{v}^-$ and $\vec{v}^+$) is an isochron (black polygons). The phase sensitivity $\vec{Z}$ is orthogonal to the isochron. A single kick of amplitude and direction $\Delta\vec{x}$ takes the shadow trajectory ($\vec{x}(\varphi)$, dark red line) from a point on the unstable manifold to a point on the stable manifold and advances the phase by $\Delta\varphi = \vec{Z}\cdot\Delta\vec{x}$.} \label{Fig:Isochrones}
\end{figure}

Measuring the frequency response will give some approximation of $\vec{Z}(\vec{x}_0)$, or rather projections of $\vec{Z}$ on the chosen perturbations. However, the response of chaotic systems is often not differentiable. 
In Fig.\ref{Fig:UPOresp}a we show the mean period as a function of system parameter $b$ in the chaotic Roessler oscillator (see Sec.\ref{sec:Roessler}). At bifurcation points where the system attractor changes non differentiably the mean period is also not differentiable. Then the limits Eq.\,\eqref{Eq:FreqShift} and Eq.\,\eqref{Eq:TimeShiftPRC} (with $\varepsilon\sim\Delta b$) may not exist and the frequency measurements can give contradicting results for different $\varepsilon$. For uniformly hyperbolic chaos, on the other hand, linear response has been proven \cite{ruelle1997differentiation}. In this class of systems the stable, unstable and neutrally stable manifolds intersect in each point $\vec{x}_0$ of the attractor and are nowhere tangential. The tangent space at $\vec{x}_0$ is spanned by the co-variant Lyapunov vectors $\vec{v}^{(k)}(\vec{x}_0)$ in the stable, unstable and neutrally stable directions. These vector fields on a hyperbolic chaotic attractor have their correspondence in the Floquet vectors on a limit cycle. In the unperturbed system a small deviation $\vec{h}(\varphi)$ from a tajectory $\vec{x}_0(\varphi)$ is evolved by the aperiodic Jacobian matrix $\textrm{J}_f(\varphi)=\textrm{J}_f(\vec{x}_0(\varphi))$
\begin{equation}
    \frac{d}{d\varphi}\vec{h} = \textrm{J}_f(\varphi)\vec{h}.
\end{equation}
In a co-moving (co-variant) base of Lyapunov vectors, the dynamics of the components $h_k$ of $\vec{h}(\varphi)=\sum_k h_k \vec{v}^{(k)}$ decouple as
\begin{equation}
    \frac{d}{d\varphi}h_k = \lambda^{(k)}(\varphi)h_k.
\end{equation}
The $\lambda^{(k)}$ are local Lyapunv exponents and the averages $\Lambda^{(k)}=\langle\lambda^{(k)}\rangle_\varphi$ are the Lyapunov exponents on the system attractor. If $\vec{f}=\vec{f}(\vec{x})$ is time independent, one Lyapunov exponent $\Lambda^{(0)}=\lambda^{(0)}=0$ is zero, globally and locally, and the corresponding Lyapunov vectorfield is $\vec{v}^{(0)}=\vec{f}$. Shifts $h_0$ in the direction of $\vec{f}$ result in time shifts which do not grow or decay. The vector field $\vec{Z}(\vec{x}_0)$ which is orthogonal to the Lyapunov vectors in the stable and unstable directions and is normalized to $\vec{Z}\cdot\vec{f}=1$ is the obvious generalization of the phase sensitivity function to chaotic oscillators (see Fig.\ref{Fig:Isochrones}). However, any perturbation with components in the unstable directions will lead to an exponentially growing deviation of a perturbed trajectory from the unperturbed trajectory. The distance between these trajectories is not bounded as $O(\varepsilon)$, they have no well defined phase relationship $\varphi(t)$ and shifts in any geometric angle-like phase due to pulsed perturbations are not asymptotically constant. 
Nevertheless, $\vec{Z}(\vec{x}_0)$ does have all necessary properties for a phase sensitivity function for one particular perturbed trajectory, which depends on the perturbation $\vec{p}(\vec{x},t)$, and shadows the unperturbed trajectory $\vec{x}_0(\varphi)$. \\ \\
{\bf Phase response conjecture for shadowing trajectories:}
Given a trajectory ${\vec x}_0(\varphi)$ on a uniformly hyperbolic invariant set 
of a flow generated by a dynamics $d{\vec x}_0/d\varphi = \vec{f}({\vec x}_0)$, and without any other continuous symmetries than time-shift invariance, for any sufficiently small perturbation $\varepsilon \vec{p}({\vec x},t)$ of arbitrary but finite duration, i.e. $|\vec{p}({\vec x},t)|=0$ for $t\notin[t_0,t_0+\tau]$, there exists a unique time isomorphism $\varphi=\varphi(t)$ with $\varphi(t_0)=t_0$ and a unique $\varepsilon$-close trajectory ${\vec x}_\varepsilon(t)$ such that $d{\vec x}_\varepsilon/dt = \vec{f}({\vec x}_\varepsilon) + \varepsilon \vec{p}({\vec x}_\varepsilon,t)$ holds exactly and $\lim_{t\to\pm\infty} |{\vec x}_\varepsilon(t)-{\vec x}_0(\varphi(t))|=0$. The time derivative of $\varphi$ in linear order of $\varepsilon$ is given by 
\begin{equation}\label{Eq:ShadowPhaseDynamics}
	\dot\varphi = 1 + \varepsilon \vec Z({\vec x}_0(\varphi))\cdot \vec{p}({\vec x}_0(\varphi),t)
\end{equation}
where the phase sensitivity function $\vec Z({\vec x}_0)$ is the unique vector field orthogonal to the stable and unstable manifolds at ${\vec x}_0$ and normalized to $\vec Z({\vec x}_0)\cdot \vec{f}({\vec x}_0) = 1$. $\Box$
\\ \\
Equation \eqref{Eq:ShadowPhaseDynamics}
defines to the linear order the instantaneous time-shift of the
shadowing trajectory relative to the unperturbed trajectory for arbitrary perturbations. After the perturbation is switched off the
shadowing trajectory $\vec{x}_\varepsilon$ will converge to the
unperturbed trajectory with an accumulated asymptotic phase 
shift $\Delta\varphi=\varphi(t)-t$. A mathematical proof of the existence of a shadowing trajectory for flows and equivalence of Lipschitz boundedness of the shadow to structural stability was given in \cite{pilyugin1997shadowing,pilyugin2010lipschitz}. In our conjecture, by imposing the boundary condition $\varphi(t_0)=t_0$ and requiring asymptotic convergence of the shadow to the unperturbed trajectory in both temporal directions the isomorphism $\varphi=\varphi(t)$ and the shadow ${\vec x}_\varepsilon(t)$ are defined uniquely. Moreover, using co-variant Lyapunov vectors \cite{ginelli2007characterizing}, the phase and the shadowing trajectory can be constructed explicitly in linear order of $\varepsilon$.
The conjecture is also valid for structurally stable invariant sets of non-hyperbolic dynamics, i.e. unstable periodic orbits (UPOs) embedded into a non-hyperbolic chaotic attractor. Phase sensitivity of UPOs has been used in \cite{pikovsky1997UPOs} to study chaotic phase synchronization. In Fig.\ref{Fig:UPOresp}b we demonstrate linear phase response by constructing the shadow of the period-1 UPO in the chaotic Roessler oscillator under periodic forcing. We chose a forcing amplitude and frequency such that the shadowing trajectory is synchronized to the forcing.
Equation \eqref{Eq:ShadowPhaseDynamics}
has the same significance as Eq.\,\eqref{Eq:LCPhaseResponse} for periodic oscillators; it is a nonlinear equation for the phase dynamics based on linear response theory, expressing the effect of a perturbation as a product of a phase sensitivity function and the perturbation itself. This makes it, for instance, possible to use linear methods to construct perturbations that optimize the response for some purpose \cite{kiss2007engineering,zlotnik2013optimal}. Using the method of linear least squares from Sec.\ref{sec:geometric} it is possible to construct a differentiable geometric phase $\vartheta_\sigma(\vec{x})$ which approximates $\vec{Z}$ on the attractor as $\vec{\nabla}\vartheta_\sigma \approx \omega_0\vec{Z}(\vec{x})$.
\subsection{Construction of the shadow trajectory}\label{sec:shadow}
Let us consider a solution $\vec{x}_0(\varphi)$ of an autonomous dynamics Eq.\,\eqref{Eq:x0_Dyn} on a hyperbolic attractor with Jacobian matrix $(\textrm{J}_f)_{ij} = \partial f_i/\partial x_j$ and a small deviation $\vec{h}(\varphi) = \sum_k \left(\vec{u}^{(k)}\cdot \vec{h}\right)\vec{v}^{(k)}=\sum_k h_k\vec{v}^{(k)}$ where the $\vec{v}^{(k)}=\vec{v}^{(k)}\left(\vec{x}_0\right)$ and the \mbox{$\vec{u}^{(k)}=\vec{u}^{(k)}\left(\vec{x}_0\right)$} are co-moving bases of biorthonormal Lyapunov vectors and co-vectors following the equations
\begin{eqnarray}
	\frac{d}{d\varphi} \vec{v}^{(k)} &=& \left[\textrm{J}_f - \lambda^{(k)}\right]\vec{v}^{(k)} \label{Eq:LyapunovEvolution} \\
	\frac{d}{d\varphi} \vec{u}^{(k)} &=& -\left[\textrm{J}_f^\top - \lambda^{(k)}\right]\vec{u}^{(k)}. \label{Eq:AdjointEvolution}
\end{eqnarray}
The $\lambda^{(k)}=\lambda^{(k)}(\varphi)\in\mathbb{R}$ are local Lyapunov exponents of the system. Equations (\ref{Eq:LyapunovEvolution},\ref{Eq:AdjointEvolution}) conserve biorthonormality $\vec{u}^{(k)}\cdot\vec{v}^{(l)}=\delta_{kl}$ along a trajectory while the  $\lambda^{(k)}$ on average compensate for the expansion or contraction in the directions of the Lyapunov vectors \cite{ginelli2007characterizing}. Equation \eqref{Eq:LyapunovEvolution} for $\lambda^{(0)}=0$ is trivially solved by $\vec{v}^{(0)}=\vec{f}$, the Lyapunov vector field corresponding to the neutrally stable direction of the flow. We now consider the evolution of $\vec{x}_\varepsilon(t) = \vec{x}_0(\varphi(t)) + \vec{h}(\varphi(t))$ in a perturbed system
\begin{equation}
	\frac{d}{dt} \vec{x}_\varepsilon = \vec{f}\left(\vec{x}_\varepsilon\right) + \varepsilon \vec{p}\left(\vec{x}_\varepsilon, t\right).
\end{equation}
Here we have introduced the isomorphism $\varphi=\varphi(t)$. 
To the linear order of $\varepsilon$ and $|\vec{h}|$ we have
\begin{eqnarray}\label{Eq:PerurbationEvolution}
	\frac{d}{d\varphi} \vec{h} &=& \frac{dt}{d\varphi} \frac{d}{dt}\vec{x}_\varepsilon - \frac{d}{d\varphi}\vec{x}_0 
					= \frac{dt}{d\varphi} \left(\vec{f} + \textrm{J}_f\vec{h} + \varepsilon\vec{p}\right) - \vec{f} \nonumber \\
					&=& \left(\frac{dt}{d\varphi}-1\right)\vec{f} + \textrm{J}_f\vec{h} + \varepsilon\vec{p}. 
\end{eqnarray}
Multiplying
Eq.\,\eqref{Eq:PerurbationEvolution} by $\vec{u}^{(k)}$ from the left, using biorthonormality,
$\frac{d}{d\varphi}\vec{h}=\sum_k h_k \frac{d}{d\varphi} \vec v^{(k)} +  \vec v^{(k)} \frac{d}{d\varphi}h_k$  and Eq.\,\eqref{Eq:LyapunovEvolution} we obtain
\begin{equation}\label{Eq:ShadowEvolution}
	\frac{dh_k}{d\varphi} = \lambda^{(k)}h_k + \varepsilon \vec{u}^{(k)}\cdot\vec{p}, \qquad \textrm{for } k\ne 0
\end{equation}
and 
\begin{equation}\label{Eq:dh0dt}
	\frac{dh_0}{dt} = \frac{dt}{d\varphi}-1 + \varepsilon \vec{u}^{(0)}\cdot \vec{p}, \qquad \textrm{for }k=0.
\end{equation}
For the correct isomorphism $t=t(
\varphi)$ the perturbed trajectory $
\vec{x}(t(\varphi))$ is always contained in the subspace spanned by the stable and unstable directions at $\vec{x}_0(\varphi)$, i.e. $dh_0/dt=0$. Therefore
\begin{equation}\label{Eq:dtdvarphi}
	\frac{dt}{d\varphi} = 1 - \varepsilon \vec{u}^{(0)}\cdot \vec{p}.
\end{equation}
Let the perturbation be of finite but arbitrary long duration, i.e. $|\vec{p}(\vec{x},t)| = 0$ for $t\notin [t_0,t_0+\tau]$. Then the sufficient conditions for convergence of the perturbed trajectory to the unperturbed trajectory for perturbations of any form are $h_0=0$, $h_k(t_0)=0$ for $\Lambda^{(k)}<0$ and $h_k(t_0+\tau)=0$ for $\Lambda^{(k)}>0$. In other words, the perturbed trajectory is fully contained in the unstable manifold at the beginning of the perturbation, hence convergence for $t\to -\infty$, and fully contained in the stable manifold at the end of the perturbation such that the shadow converges to the unperturbed trajectory for $t\to\infty$ (see Fig.\ref{Fig:Isochrones}). Using these as initial and final conditions we can integrate Eqs.\,(\ref{Eq:ShadowEvolution},\ref{Eq:dtdvarphi}) forward in time, beginning at $t_0$ for the components of $\vec{h}$ in the stable directions and backward in time beginning at $t_0+\tau$ for the components in the unstable directions. Equation \eqref{Eq:ShadowEvolution} being linear, the distance $|\vec{h}|$ of the perturbed trajectory is always of order $\varepsilon$ on uniformly hyperbolic invariant sets, i.e. when the dynamics in the stable and unstable directions is uniformly contracting or expanding. Furthermore denoting $\vec{Z} = \vec{u}^{(0)}$ Eq.\,\eqref{Eq:dtdvarphi} is identified as Eq.\,\eqref{Eq:TimeResponse} and can in linear order of $\varepsilon$ be rewritten  as Eq.\,\eqref{Eq:ShadowPhaseDynamics}. It is, however, easier to integrate Eq.\,\eqref{Eq:dtdvarphi} when $\vec{x}_0(\varphi)$ is given at discrete time points $\varphi_i$. Using $\varphi(t_0) = t_0$ as initial condition makes the time isomorphism unique. The phase sensitivity function $\vec{Z}(\vec{x}_0)$ is the unique vector field solving the adjoint equation $d\vec{Z}/d\varphi = -\textrm{J}_f^\top \vec{Z}$, i.e. Eq.\,\eqref{Eq:AdjointEvolution} for $k=0$, on the hyperbolic attractor normalized to $\vec{Z}\cdot\vec{f}=1$. 
\subsection{Discussion}\label{sec:discuss}
This method of generating a shadowing trajectory is of equivalent accuracy as a recently proposed linear least squares method \cite{wang2014least} which, however, cannot reproduce the correct time isomorphism or the phase sensitivity.
It is reasonable to assume linear response theory is valid for perturbations $\varepsilon\vec{p}$ that lead to small distances $|\vec{h}|$ in Eq.\,\eqref{Eq:ShadowEvolution}. The more the stable and unstable directions $\vec{v}^{(k)}$ align, the larger the Lyapunov co-vectors $\vec{u}^{(k)}$ become. This puts practical limits on the perturbation strength. If $|\vec{h}|\le h_\textrm{max}$ for a given perturbation $\varepsilon\vec{p}$, then $h_\textrm{max}$ depends linearly on $\varepsilon$. In particular for non-uniformly hyperbolic systems $h_\textrm{max}$ may occasionally become very large even for small $\varepsilon$. Furthermore, while the trajectory $\vec{x}_\varepsilon(t)$ is an exact solution of the perturbed system, it may however not be a typical solution, i.e. time averages are not necessarily equal to averages with respect to the natural invariant measure on the perturbed attractor. If the chaotic attractor is not structurally stable then in the vicinity of a larger bifurcation, e.g. a periodic window, the attractor, and thus the oscillation period, can change discontinuously in response to the perturbation (see Fig.\,\ref{Fig:UPOresp}a). The measured frequency response Eqs.\,(\ref{Eq:FreqShift},\ref{Eq:TimeShiftPRC}) may only approximately be predicted by the projection of the perturbation on $\vec{Z}$, i.e. by the averaged response of the UPOs embedded in the chaotic attractor, which are not close to a bifurcation. 
\\ \\
Although many analytic results are valid for hyperbolic systems, physical examples of hyperbolic chaotic flows are rare \cite{kuznetsov2007autonomous}. On the other hand, the algorithm \cite{ginelli2007characterizing} for the numerical determination of the Lyapunov vectors is quite robust against occasional near tangencies of stable and unstable manifolds along the trajectories on the chaotic attractor. Even for non-hyperbolic systems such as the Roessler system for small enough perturbations one can construct shadowing trajectories which remain close to an unperturbed trajectory for periods of time longer than expected from the largest rate of divergence given by the largest Lyapunov exponent. In Sections \ref{sec:Gaspard} and \ref{sec:Roessler} we present examples of non-hyperbolic chaotic oscillators where our method can reliably predict the frequency response. In Section \ref{sec:Kuzn} we show that our method works with a known example of hyperbolic chaotic oscillations and in \ref{Sec:Lorenz} we discuss why our method works poorly in the non-hyperbolic Lorenz system.
\section{Examples}\label{Sec:Examples}
\begin{figure}[h!]
\setlength{\unitlength}{1cm}
\begin{picture}(3.4,3.4)
\put(-0.5,0){\includegraphics[width=4cm]{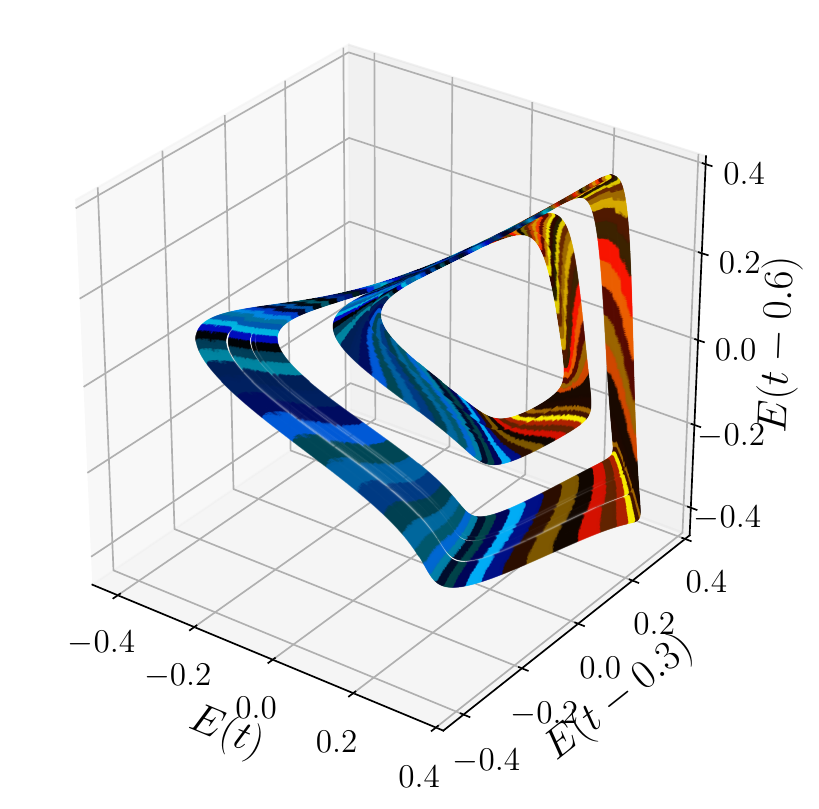}}
\put(-0.8,3.1){\bf (a)}
\end{picture}
\begin{picture}(3.4,3.4)
\put(0,0){\includegraphics[width=4cm]{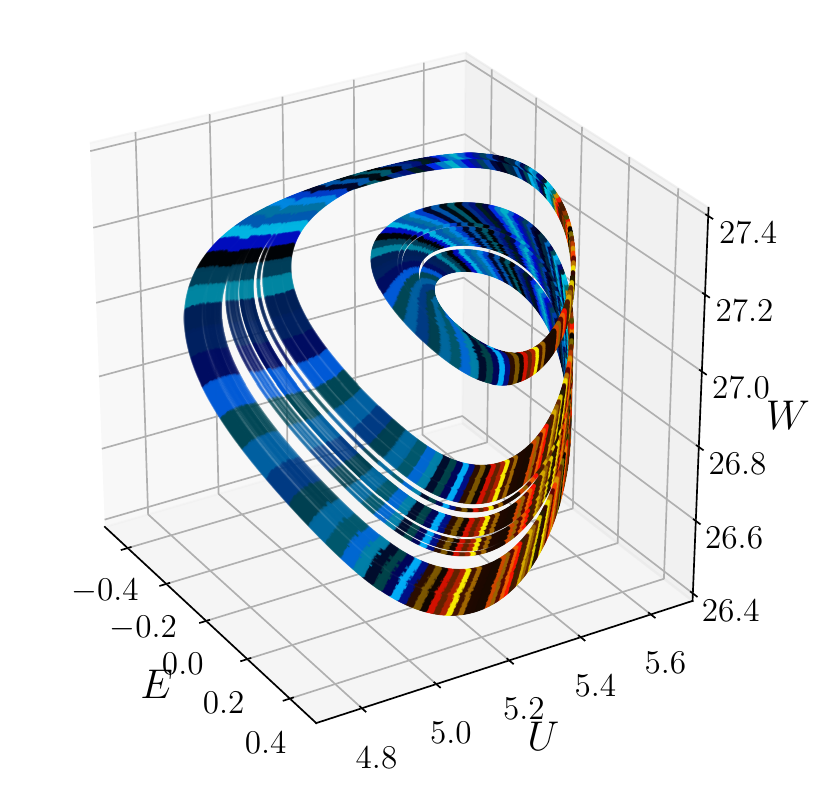}}
\put(0,3.1){\bf (b)}
\end{picture}
\begin{picture}(3.4,3.4)
\put(-0.2,0){\includegraphics[width=3.8cm]{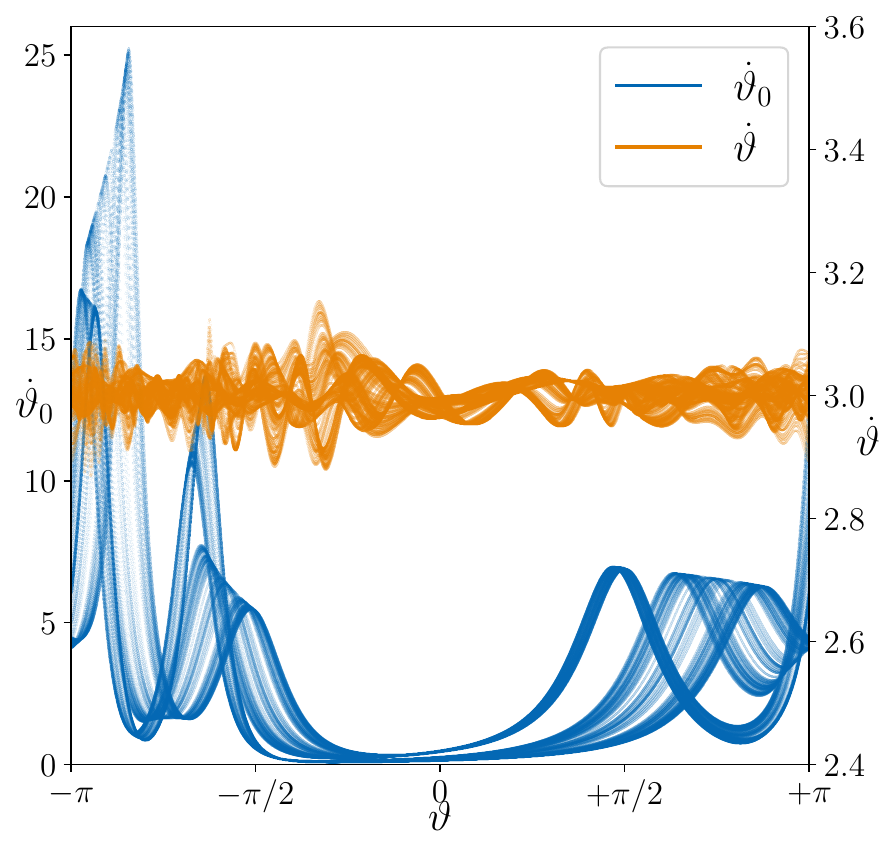}}
\put(-0.8,3.1){\bf (c)}
\end{picture}
\begin{picture}(3.4,3.4)
\put(0,0){\includegraphics[width=3.8cm]{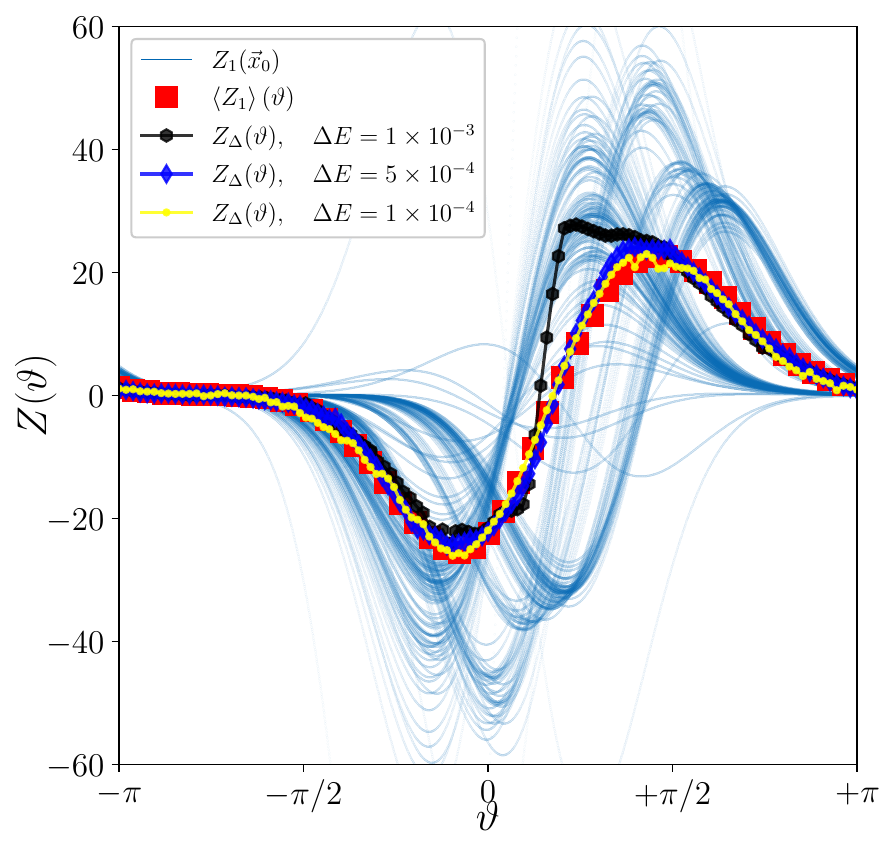}}
\put(0,3.1){\bf (d)}
\end{picture}
\begin{picture}(3.4,3.4)
\put(-0.4,0){\includegraphics[height=3.6cm]{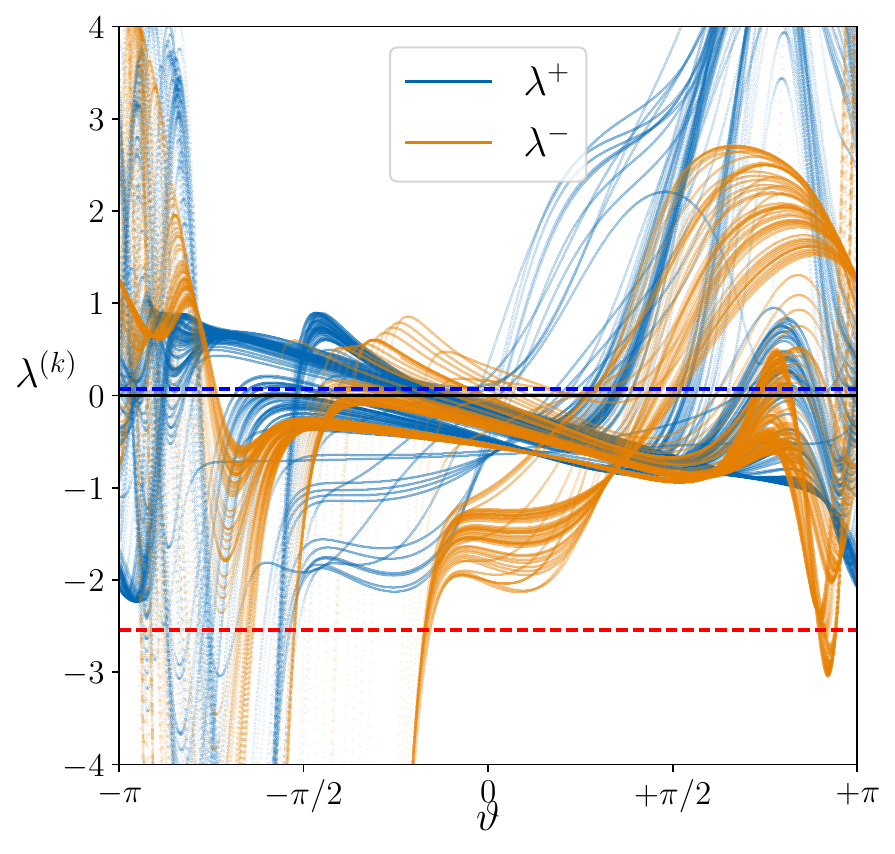}}
\put(-0.8,3.1){\bf (e)}
\end{picture}
\begin{picture}(3.4,3.4)
\put(0,0){\includegraphics[height=3.6cm]{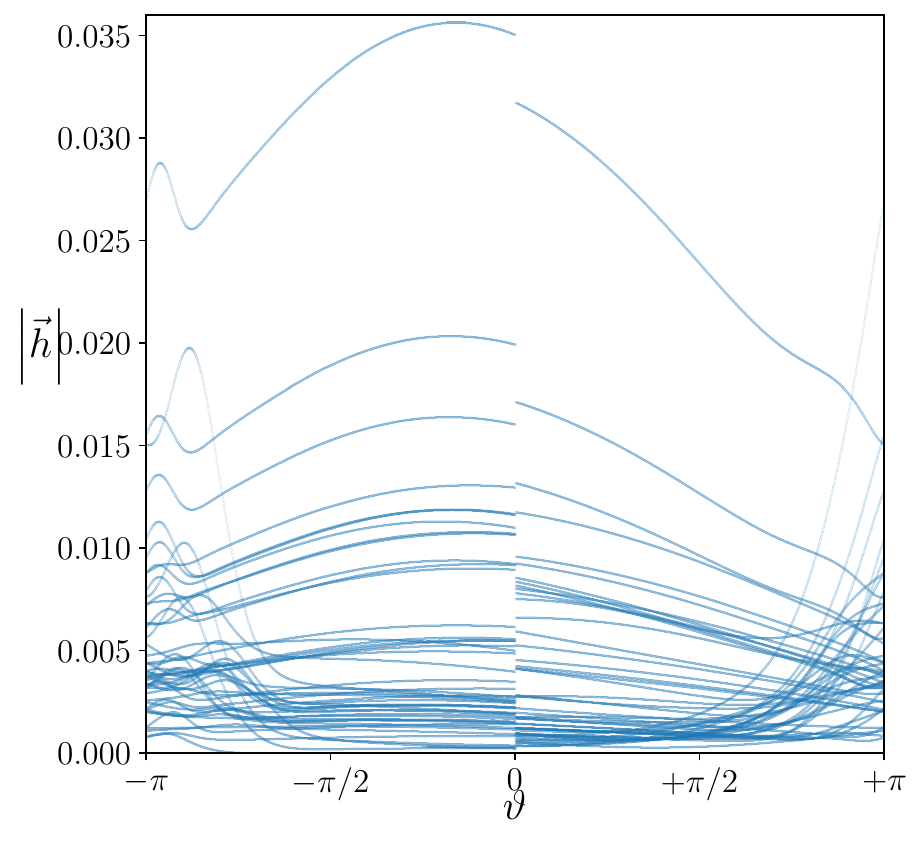}}
\put(0,3.1){\bf (f)}
\end{picture}
\caption{Numerical integration of chaotic electro-chemical oscillator model Eqs.\,\eqref{Eq:GaspardOsc}-\eqref{Eq:GaspardNonlin} over $400$ time units with $d\varphi=1\times 10^{-3}$. Transients for the convergence of Lyapunov vectors have been discarded. (a) chaotic attractor in the time-delay embedding $(x,y,z) = (E(t),E(t-0.3),E(t-0.6))$ Color coded are small intervals of the optimized phase $\vartheta=\vartheta_\sigma$. Blue shades signify regions of positive PRC and red hues negative values. (b) Chaotic attractor in the original dynamic variables $(E,U,W)$. The color code of the phase intervals is the same as for the corresponding points in (a). In (c) we show the velocity of the geometric proto-phase $\vartheta_0$ (blue lines, left axis) and compare them to the velocity of the optimized phase (orange lines, right axis) with much smaller standard deviation (3.67 vs. 0.03). Both phase velocities are shown as functions of the optimized geometric phase. In subfig. (d) we compare the component of $\vec{Z}({\vec{x}_0})$ in the $E$ direction obtained by the method of Lyapunov vectors (light blue lines) and their average at constant angle $\vartheta$ (large red squares) with frequency response curves obtained from kicking the oscillator in the $E$ direction every time the Poincare section $P_\vartheta$ is crossed after completing one rotation. Up to a kick strength of $\Delta E \le 5\times 10^{-4}$ the curves follow the theoretical prediction via the method of Lyapunov vectors. (e) Local Lyapunov exponents $\lambda^{(+)}$ and $\lambda^{(-)}$ for the Lyapunov vectors in the unstable and stable directions.  Both have large deviations in the positive and negative directions, but $\Lambda^{(+)}=\langle\lambda^{(+)}\rangle = 0.07$ (blue dashed line) and $\Lambda^{(+)}=\langle\lambda^{(+)}\rangle = -2.5$ (red dashed line) are rather small. Finally in (f) we show the distance $|\vec{h}|$ of the shadow trajectory which is kicked at optimized geometric phase $\vartheta=0$ with strength $\Delta E = 1\times 10^{-4}$. For larger values of $\Delta E$ the distance of the shadow in linear approximation would increase proportionally.
}\label{Fig:Gaspard}
\end{figure}
\subsection{Electrochemical oscillations}\label{sec:Gaspard}
As an example we consider current oscillations during the electro-dissolution of a metal in an acidic environment. A mathematical model for such electro-chemical oscillations, which exhibits a period doubling route to non-hyperbolic chaos, was developed in \cite{koper1992modeling} and used in \cite{kiss2001phase} to reproduce in simulations the experimentally observed chaotic current oscillations through a nickel electrode in sulfuric acid. After an appropriate re-scaling we obtain
\begin{eqnarray}\label{Eq:GaspardOsc}
    \dot E &=& \frac{V_{a}-(E+36)}{R_s}-6g(E)U \\
    \dot U &=& -1.25\sqrt{d}g(E)U + 2d(\frac{1}{15}W+\frac{40}{3}-U) \\
    \dot W &=& 1.6 d (15U-3W)
\end{eqnarray}
with nonlinearity
\begin{equation}\label{Eq:GaspardNonlin}
    g(E) = 2.5 e^{-(E+1)^2} + 0.01 e^{\frac{1}{2}(E+6)}.
\end{equation}
The applied Voltage $V_a$ and the electrode potential drop $E$ can be measured and controlled. At the parameters $V_a=36.7380$, $R_s=0.02$ and $d=0.119$ the system attractor has developed two chaotic bands around an unstable period-two orbit. The applied voltage $V_a$ needs to be controlled precisely since the region of chaotic oscillations in parameter space is very small. Only the phase sensitivity in the $E$ component is of experimental interest since $U$ and $W$ quantify a gradient of chemical concentrations in the solution (double layer approximation) and cannot be measured. However, for the computation of the Lyapunov vectors the full knowledge of the system state, velocity and Jacobian are assumed.
We have calculated the Lyapunov exponents on the chaotic attractor as $(\Lambda^{(0)},\Lambda^{(+)},\Lambda^{(-)})=(0,0.07,-2.5)$. For Lyapunov vectors $\vec{v}^{(\pm)}$ of unit length, the local Lyapunov exponents $\lambda^{(\pm)}$ exhibit large excursions to both positive and negative values (Fig.\ref{Fig:Gaspard}e). As a consequence the deviations of a shadowing trajectory in Eq.\,\eqref{Eq:ShadowEvolution} can become quite large, even for small perturbations.
\\ \\
We define $x=E$, $y=E(t-0.3)$ and $z=E(t-0.6)$. A geometric proto-phase $\vartheta_0$ with positive phase velocity and an amplitude $R$ can be defined via $x=R\cos\vartheta_0$ and $y=R\sin\vartheta_0$.
The phase $\vartheta = \vartheta_{\sigma}(\vartheta_0,R,z)$ that we want to optimize is expanded into
\begin{equation}
    \vartheta_{\sigma} = \vartheta_0 + \sum_{k=0}^{6}\sum_{l=0}^{3}\sum_{m=0}^{3} \sigma^\pm_{klm}q^\pm_{klm}
\end{equation}
with 
\begin{equation}
    q^+_{klm} = \cos(k\vartheta_0)R^lz^m, \qquad
    q^-_{klm} = \sin(k\vartheta_0)R^lz^m,
\end{equation}
$\sigma^+_{000}=0$ and $\sigma^-_{0lm}=0$.
The choice of the cutoff values for the Fourier harmonics $k$ and polynomial orders $l$, $m$ depends on the particular geometry of a system. Lower values avoid over-fitting with large deviations at points that are not on the attractor, whereas larger values can give better results for the points on the attractor.
Since the stable and unstable directions in the time delayed coordinates are not known we only use Eq.\,\eqref{Eq:WLLS} with $l=0$ and $\vec{f}=\frac{d}{d\varphi}(x,y,z)$ to minimize the variance of the deviations $\eta_0$ in
\begin{equation}
    \dot\vartheta_\sigma = \dot\vartheta_0 + \sum_k \sigma_k \dot q_k = \omega_0 + \eta_0.
\end{equation}
The delay embedding of the chaotic attractor with color-coded optimized phase is shown in Fig.\ref{Fig:Gaspard}a. The velocity of the proto-phase and of the optimized phase as functions of $\vartheta$ are shown in Fig.\ref{Fig:Gaspard}c. In both cases the mean phase velocity is $\omega_0=3.001$ but the standard deviation of the optimized phase velocity is at $0.03$ within 1\% of $\omega_0$.
\\ \\
Next we perform a series of perturbation experiments. A small delta kick in the applied potential $V_a$ is executed after each full rotation when the system crosses the Poincare section $P_{\vartheta}$ at a given optimized geometric phase $\vartheta$ in the delay coordinates. The measured shift in the average period according to Eq.\,\eqref{Eq:TimeShiftPRC} gives the PRC $Z_\Delta(\vartheta)$ (Fig.\ref{Fig:Gaspard}d, small markers). This PRC can be compared with the components $Z_1$ of $\vec{Z}(\vartheta)$ in the $E$ direction. Here $\vec{Z}=\vec{Z}(\vec{x}_0)$ is calculated numerically from the co-variant Lyapunov vectors \cite{ginelli2007characterizing}.
As a function of $\vartheta(\vec{x}_0)$ the values of $Z_1(\vec{x}_0)$ form a family of curves, shown as thin blue lines in Fig.\ref{Fig:Gaspard}d.
An average response $\langle Z_1\rangle(\vartheta)$ is calculated via a narrow Gaussian filtering of the data points $Z_1(\vartheta)$ (red squares). For this chaotic oscillator our linear frequency response theory predicts the measured PRC $Z_\Delta(\vartheta)$ very well. However the strength of the delta kicks must be very small $\left(\Delta E < 5\times 10^{-4}\right)$ in order to approximately retain the structure of the chaotic attractor, and even smaller $\Delta E\approx 1\times 10^{-4}$ for a shadow trajectory which in linear order of the perturbation stays within an acceptable small distance to the unperturbed trajectory (Fig.\ref{Fig:Gaspard}f).
\begin{figure}[!t]
\setlength{\unitlength}{1cm}
\begin{picture}(4.2,4.2)
\put(0,0){\includegraphics[height=4.4cm]{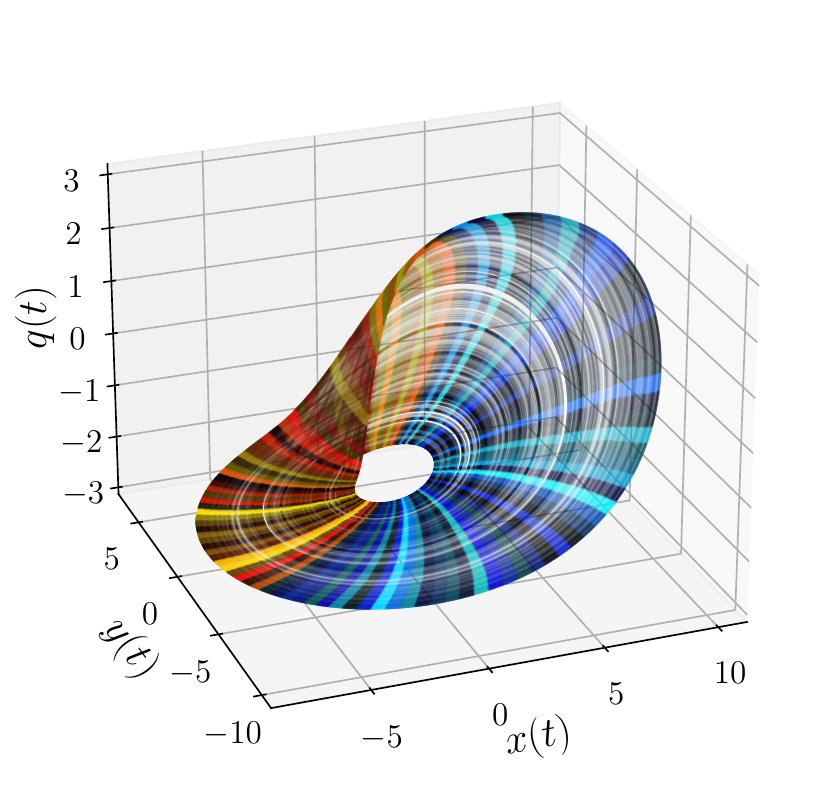}}
\put(0,3.8){\bf (a)}
\end{picture}
\begin{picture}(4.2,4.2)
\put(-0.1,0){\includegraphics[height=4.2cm]{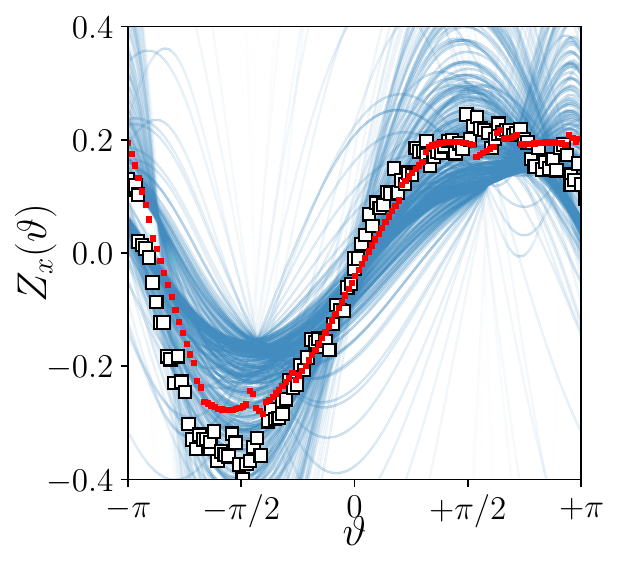}}
\put(0,3.8){\bf (b)}
\end{picture}
\begin{picture}(4.2,4.2)
\put(-0.3,0){\includegraphics[height=4.2cm]{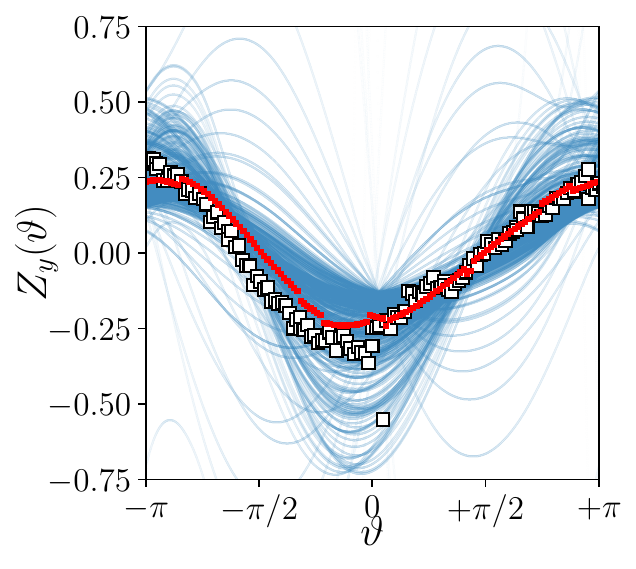}}
\put(0,3.75){\bf (c)}
\end{picture}
\begin{picture}(4.2,4.2)
\put(-0.1,0){\includegraphics[height=4.2cm]{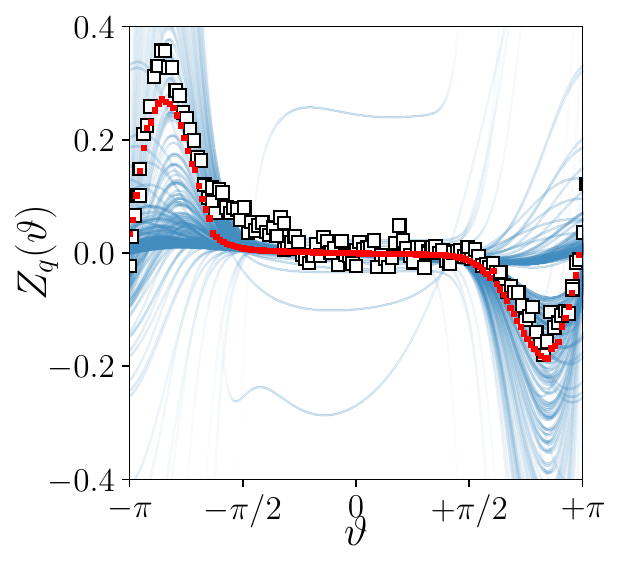}}
\put(0,3.8){\bf (d)}
\end{picture}
\caption{Frequency response in the chaotic Roessler system \eqref{Eq:RoesslerSystem}. (a) Chaotic attractor with color coded small intervals of optimized geometric phase $\vartheta(\vec{x}_0)$. Blue hues indicate negative values of $Z_x$ and red hues positive values. Panels (b-d) show the components of the phase sensitivity $\vec{Z}(\vec{x}_0)$ (thin blue lines) as a function of the optimized geometric phase $\vartheta$, a narrow Gaussian average of these values as red dots, disregarding values of $Z$ larger than three standard deviations, and (white square markers) the linear response of the oscillation period to delta kicks of strength $\varepsilon=0.05$ in the three dynamical variables (b) $x$, (c) $y$ and (d) $q$ at the crossing of a given Poincare section in the optimized geometric phase after each full rotation.}\label{Fig:Roes}
\end{figure}

\subsection{Roessler oscillator}\label{sec:Roessler}
The chaotic Roessler oscillator is often used as an example for chaotic phase synchronization \cite{rosenblum1996phase,pikovsky2003synchronization,Ott98}. Chaotic phase diffusion in the Roessler system is extremely small \cite{Ott98}, which facilitates phase synchronization in this system. Tangencies of the Lyapunov vectors occur but the directions of the co-variant Lyapunov vectors are well separated most of the time \cite{ginelli2007characterizing}. The transition to chaos occurs via period doubling with frequent bifurcations of the attractor where the response is not differentiable (Fig.\ref{Fig:UPOresp}a). However, these structural changes in the attractor can be small if the main UPOs are not close to a bifurcation. We study the chaotic Roessler oscillator with a logarithmic variable $z = \exp(q)$, effectively making additive perturbations in $q$ multiplicative in $z$, ensuring that $z$ remains positive. The dynamics in these variables is given by
\begin{eqnarray}\label{Eq:RoesslerSystem}
    \dot x &=& -y - e^{q} \\
    \dot y &=& x + a y \\
    \dot q &=& be^{-q}+(x-c) 
\end{eqnarray}
where we have used $a=0.25$, $b=0.9$ and $c=6.0$.
Unusually large values of the phase sensitivity $\vec{Z}(\varphi)$ (Lyapunov co-vector) do occur which have a strong influence on the average phase response. In the averages $\langle Z_i\rangle(\vartheta)$ we therefore disregard values of the phase sensitivity larger than three standard deviations. A protophase $\vartheta_0$ and radial distance $R$ for this system is defined as $x = R\cos\vartheta_0$ and $y=R\sin\vartheta_0$. For the optimized phase we use the same expansion and cutoff as in the previous example of the electrochemical oscillator (Sec.\ref{sec:Gaspard}). However, we determine the optimized phase using the full information of the unit length Lyapunov vectors $\vec{v}^\pm$ and the flow direction $\vec{v}^0=\vec{f}$. The resulting optimized phase $\vartheta(\vec{x})$ is then used in the perturbation experiments to measure the average shift in the rotation period in response to delta kicks at the crossing of a given Poincare section $P_\vartheta$ after each full rotation. We compare the predicted average response by the method of Lyapunov vectors to the measured response to delta kicks of strength $\varepsilon = 0.05$ in Fig.\ref{Fig:Roes}.
\subsection{Hyperbolic chaotic oscillations}\label{sec:Kuzn}
We will now demonstrate our theory at the following example of hyperbolic chaotic dynamics \cite{kuznetsov2007autonomous}
\begin{eqnarray}
	\dot x_1 &=& 2\pi y_1 + \left(1-a^2_2+\frac{1}{2}a^2_1-\frac{1}{50}a_1^4\right) x_1 + \kappa x_2 y_2 \label{Eq:FlipFlop11}\qquad\\
	\dot y_1 &=& -2\pi x_1 +  \left(1-a^2_2+\frac{1}{2}a^2_1-\frac{1}{50}a_1^4\right) y_1  \label{Eq:FlipFlop12}\\
	\dot x_2 &=& 2\pi y_2 + \left(a_1^2-1\right) x_2 + \kappa x_1 \label{Eq:FlipFlop21}\\
	\dot y_2 &=& -2\pi x_2 + \left(a_1^2-1\right) y_2. \label{Eq:FlipFlop22} 
\end{eqnarray}
\begin{figure}[!t]
\setlength{\unitlength}{1cm}
\begin{picture}(4.2,4.2)
\put(0,0){\includegraphics[height=4.2cm]{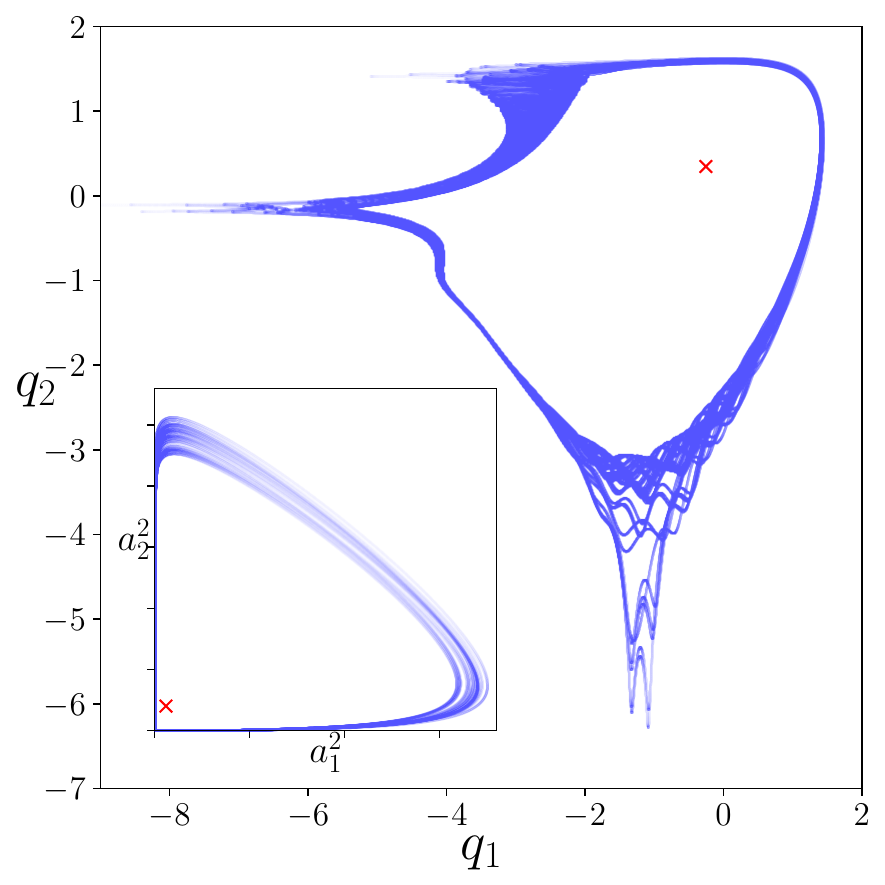}}
\put(0,3.8){\bf (a)}
\end{picture}
\begin{picture}(4.2,4.2)
\put(0,0){\includegraphics[height=4.2cm]{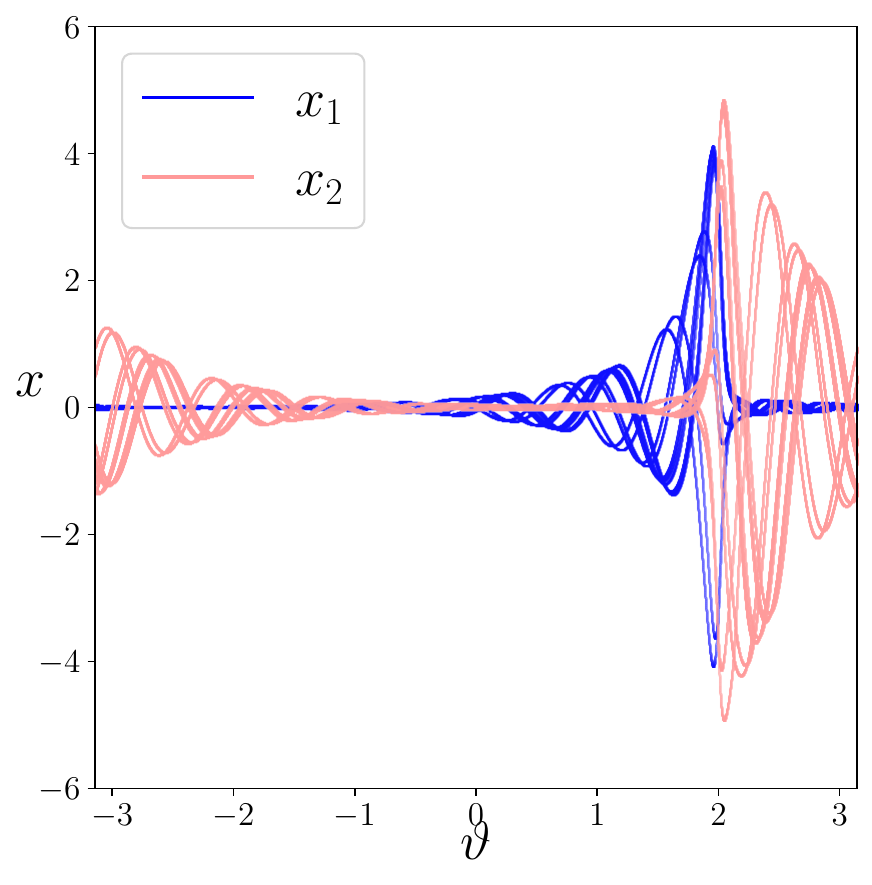}}
\put(0,3.8){\bf (b)}
\end{picture}
\begin{picture}(4.2,4.2)
\put(0.1,0){\includegraphics[height=4.2cm]{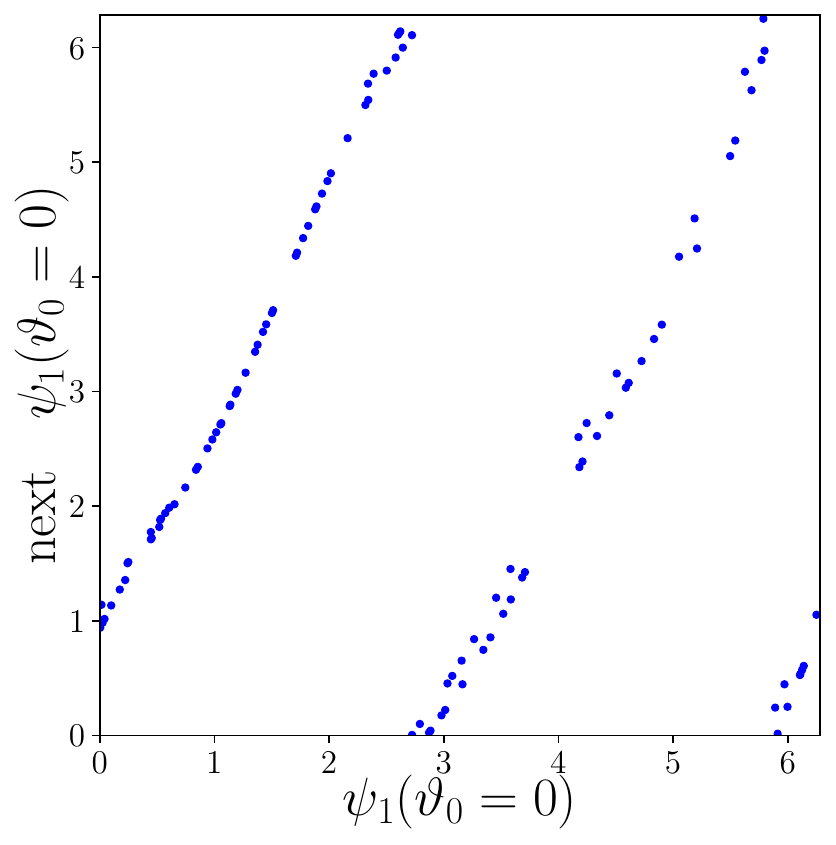}}
\put(0,3.8){\bf (c)}
\end{picture}
\begin{picture}(4.2,4.2)
\put(0,0){\includegraphics[height=4.2cm]{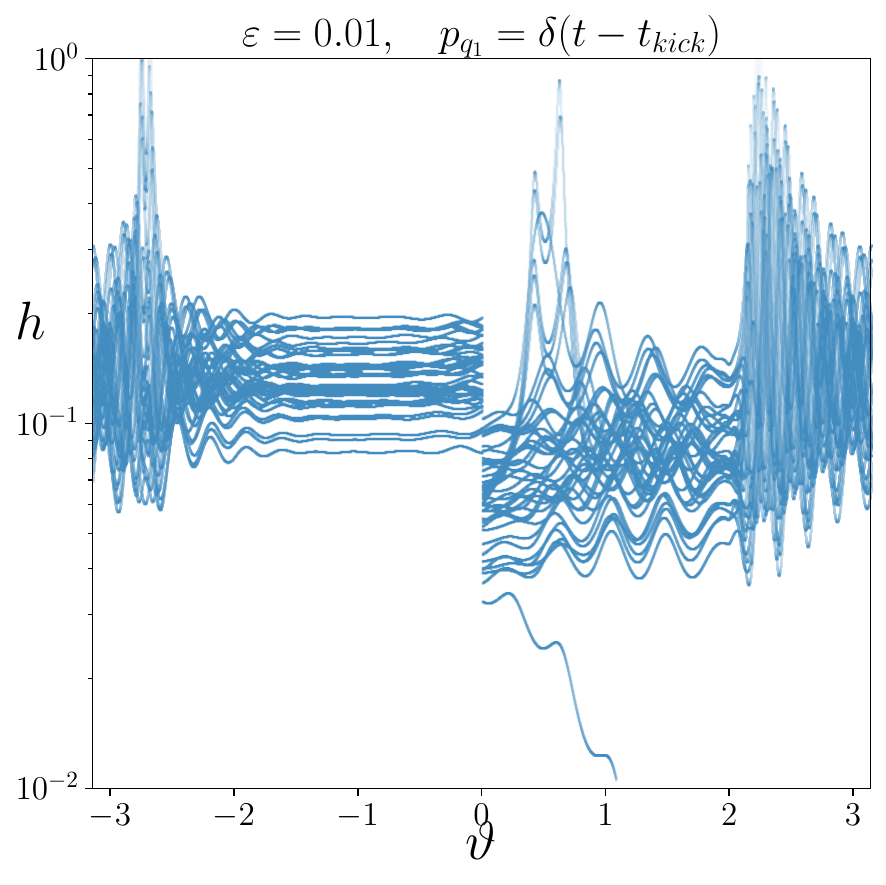}}
\put(0,3.65){\bf (d)}
\end{picture}
\begin{picture}(4.2,4.2)
\put(0,0){\includegraphics[height=4.2cm]{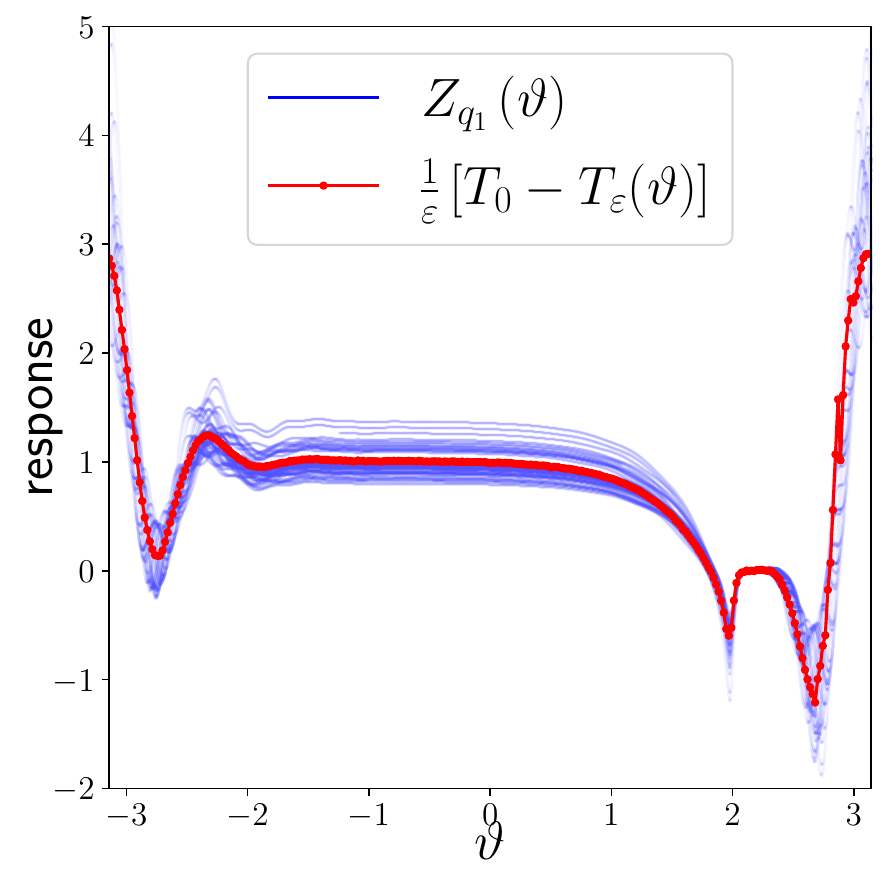}}
\put(0,3.8){\bf (e)}
\end{picture}
\begin{picture}(4.2,4.2)
\put(0,0){\includegraphics[height=4.2cm]{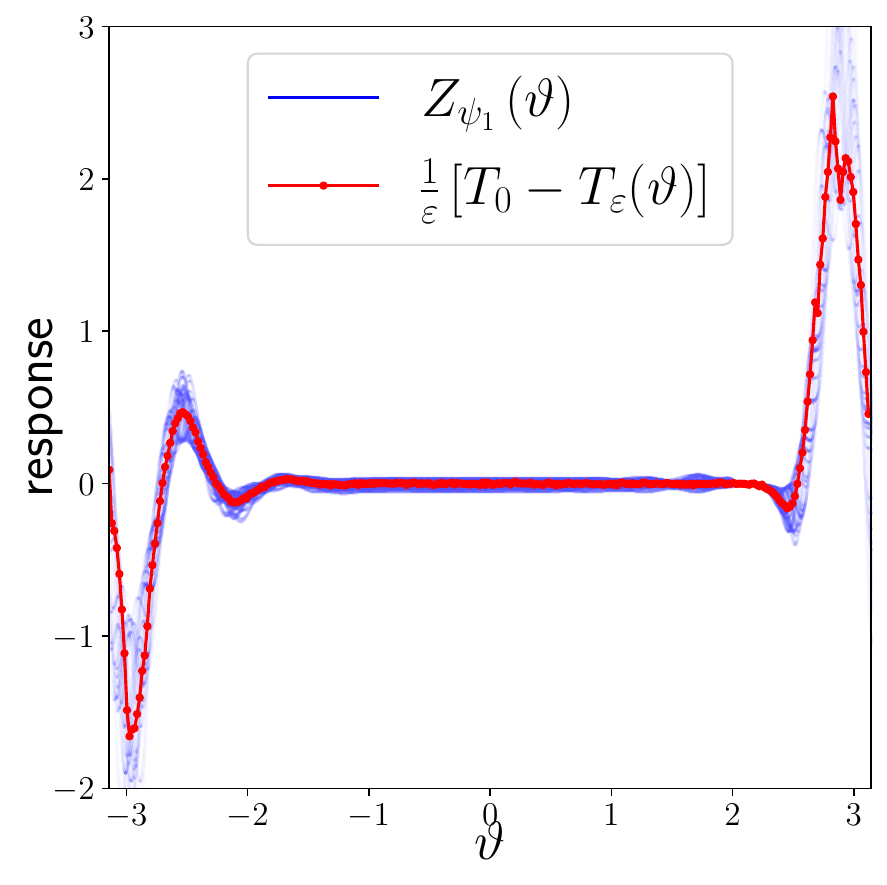}}
\put(0,3.8){\bf (f)}
\end{picture}
\caption{Hyperbolic activator-inhibitor dynamics (\ref{Eq:FlipFlop11}-\ref{Eq:FlipFlop22}) of two coupled oscillators with chaotic phase dynamics \cite{kuznetsov2007autonomous}. (a) log-amplitudes $q_i=\log a_i$ and square amplitudes $a_i^2 = x_i^2+y_i^2$ (shown in inset). (b) $x$ coordinates of the two oscillators as a function of geometric phase $\vartheta$ over 9 periods of the amplitude oscillations. (c) The Poincare map of the angle $\psi_1$ at geometric phase $\vartheta_0=0$ is an expanding circle map. (d) Distance $h=|\vec{h}|$ of the perturbed trajectory from an unperturbed trajectory for log-amplitude $\delta$-Kicks of strength $\varepsilon=0.01$ at geometric phase $\vartheta=0$. The distance after the kick is smaller than before because relaxation in the unstable directions is slower and in that direction the shadowing trajectory is by construction kicked back to the unperturbed trajectory. (e) Component $Z_{q_1}$ and (f) component $Z_{\psi_1}$ of the Lyapunov co-vector $\vec{Z} = \vec{u}^{(0)}$ (thin lines) and period response (\ref{Eq:FreqShift}) to kicking the log-amplitude $q_1$ or the angle $\psi_1$ of the first oscillator at a given geometric phase $\vartheta$ (dot markers) with $\varepsilon=0.1$.}\label{Fig:Kuz}
\end{figure}
The amplitudes $a_1^2 = x_1^2+y_1^2$ and $a_2^2=x_2^2+y_2^2$ of two oscillators are coupled via a negative feedback loop where the first oscillator acts as an activator and the second as an inhibitor leading to sequential switching between low and high amplitude oscillations. Through weak forcing with $\kappa=0.3$ the phase of the lower amplitude oscillator synchronizes to the phase of the high amplitude oscillator. By coupling the first oscillator to the second harmonics of the second oscillation via the product $\kappa x_2 y_2$ the phases of the oscillators after each round of switching are chaotic following an expanding circle map. The system's Lyapunov exponents are $\lambda \in\left\lbrace -1.34,-0.97, 0,0.09\right\rbrace$. Because the amplitudes can become very small, for numerical stability we simulate (\ref{Eq:FlipFlop11}-\ref{Eq:FlipFlop22}) using angle and log-amplitude variables $\psi$ and $q$, i.e. $x+iy=\exp(q+i\psi)$. The phase sensitivity has components $\vec{Z}=(Z_{q_1},Z_{q_2},Z_{\psi_1},Z_{\psi_2})$ in these variables corresponding to delta Kicks in the log-amplitudes and angles or $\vec{p}_q = (x,y)\delta(t-t_0)$ and $\vec{p}_\psi = (-y,x)\delta(t-t_0)$ in the original variables.
As Poincare sections we define the sets of geometric angles $\vartheta_0$ with $a_1^2-\left\langle a_1^2 \right\rangle = R\cos\vartheta_0$ and $a_2^2-\left\langle a_2^2 \right\rangle = R\sin\vartheta_0$. We re-parametrize these angles $ \vartheta_0\to\vartheta$ such that $\vartheta$ is uniformly distributed over $[0,2\pi)$. This is achieved by defining $\vartheta(\vartheta_0)$ linearly increasing with the rank of the protophases sorted over the points of the attractor. Optimization of the shapes of the Poincare sections is not necessary. In Fig.\ref{Fig:Kuz}a we show a projection of the hyperbolic chaotic attractor in the $(q_1,q_2)$ plane and $(a_1^2,a_2^2)$ in the inset. The switching dynamics can be seen in Fig.\ref{Fig:Kuz}b where $x_1$ and $x_2$ are plotted as a function of $\vartheta$. The mapping of the angle $\psi_1$ of the first oscillator from one crossing of the Poincare section $\vartheta_0=0$ to the next is shown in Fig.\ref{Fig:Kuz}c. It follows an expanding circle map. In Fig.\ref{Fig:Kuz}e,f we show the components $Z_{q_1}$ and $Z_{\psi_1}$ of the phase sensitivity function, i.e. the respective components of the Lyapunov co-vector $\vec{u}^{(0)}$ with $\vec{u}^{(0)}\cdot\vec{f}=1$, as a function of $\vartheta$, corresponding to phase shifts from perturbations in the log-amplitude or the angle of the first oscillator. Next we measured the frequency shift caused by pulsed perturbations in the log-amplitude  or the angle of the first oscillator as a function of $\vartheta$. That is, after each full oscillation of the system at a Poincare section at a given geometric phase $\vartheta$ the log-amplitude or the angle of the first oscillator was increased by $\varepsilon=0.1$ and the mean period was determined by the elapsed time between $1000$ crossings. The the frequency response function at $\vartheta$ is then calculated as
$
	Z_\Delta(\vartheta) = \frac{1}{\varepsilon}\left(T_0-T_\varepsilon\right).
$
This measure is also shown in Fig.\ref{Fig:Kuz}e,f and it traces the correpsonding components of the Lyapunov co-vector field $\vec{Z}$ very well. Finally, by integrating (\ref{Eq:ShadowEvolution}) with kicked log-amplitude $q_1$ perturbations of strength $\varepsilon=0.01$ at crossings of $\vartheta=0$ we have calculated the displacements $\vec{h}(\vartheta)$ of the shadowing trajectory and plotted $h=|\vec{h}|$ as a function of $\vartheta$ in Fig.\ref{Fig:Kuz}d. 
\begin{figure}[!t]
\setlength{\unitlength}{1cm}
\begin{picture}(4.2,4.2)
\put(0,0){\includegraphics[height=4.4cm]{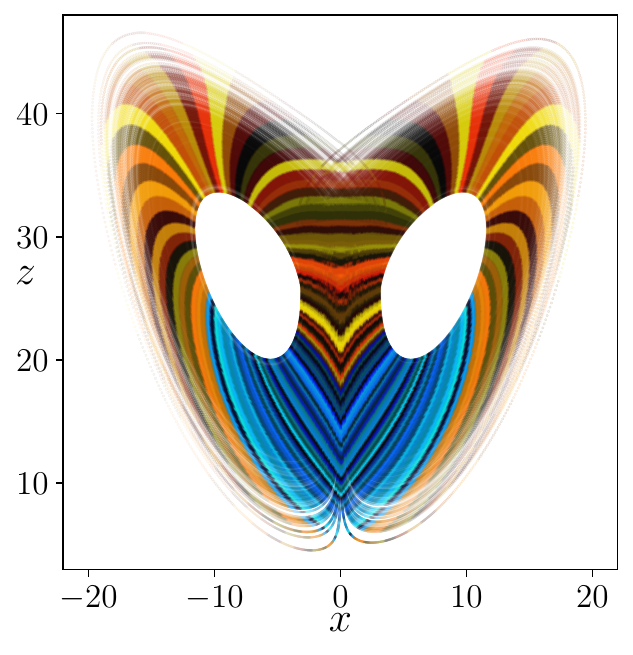}}
\put(0,3.8){\bf (a)}
\end{picture}
\begin{picture}(4.2,4.2)
\put(-0.1,0.1){\includegraphics[height=4.2cm]{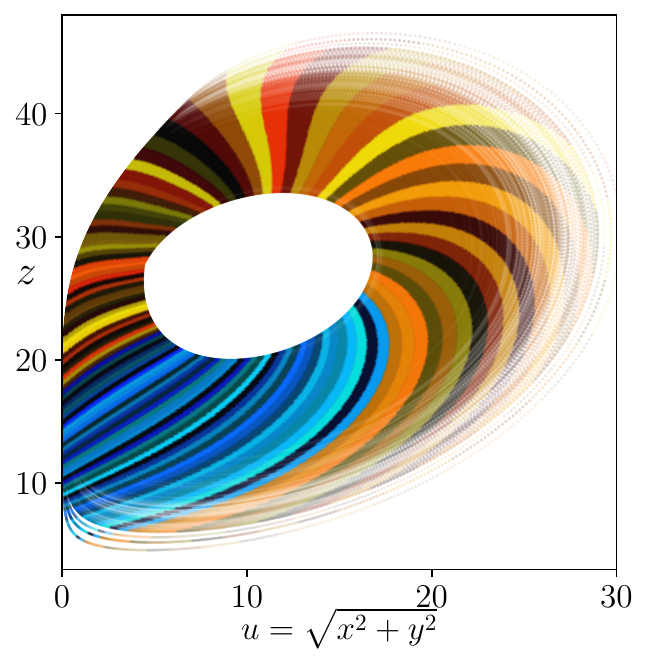}}
\put(-0.1,3.8){\bf (b)}
\end{picture}
\begin{picture}(4.2,4.2)
\put(-0.3,0){\includegraphics[height=4.2cm]{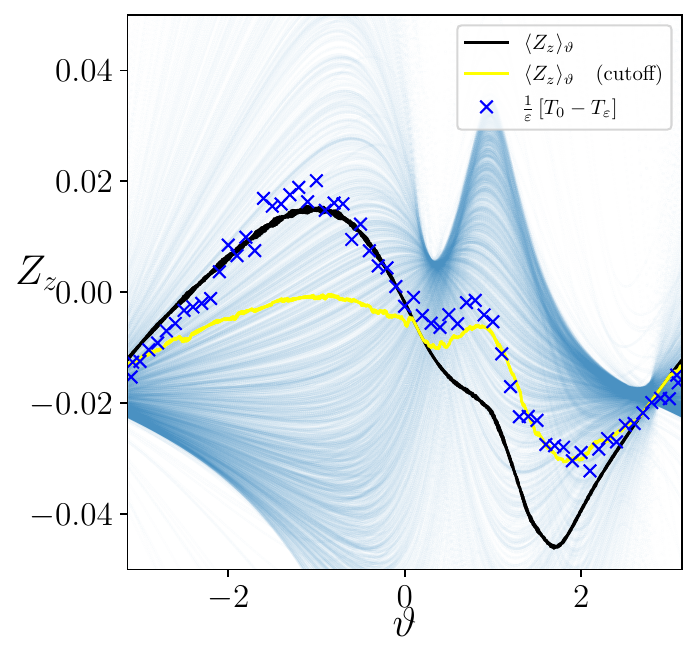}}
\put(0,3.84){\bf (c)}
\end{picture}
\begin{picture}(4.2,4.2)
\put(-0.1,0){\includegraphics[height=4.2cm]{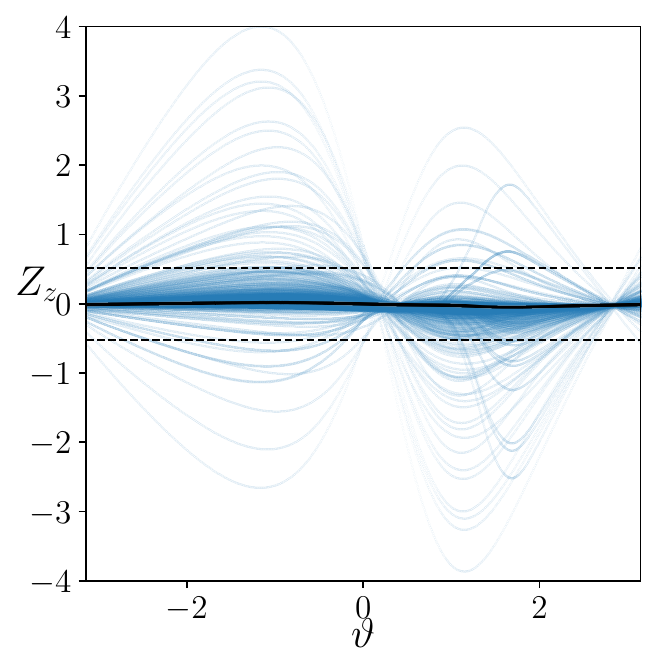}}
\put(-0.1,3.8){\bf (d)}
\end{picture}
\caption{Frequency response in the chaotic Lorenz system \eqref{Eq:LorenzSystem}. (a) Projection of the chaotic attractor to the coordinates $(x,z)$ with color coded small intervals of optimized geometric phase $\vartheta$. Blue hues indicate negative values of $Z_z$ and red hues positive values. (b) Projection of the chaotic attractor to coordinates $(\sqrt{x^2+y^2},z)$ with the same intervals of optimized geometric phase. (c) Component $Z_z$ of the phase sensitivity (family of thin blue lines) as a function of $\vartheta$, average value of $\langle Z_z\rangle_\vartheta$ (black line), average restricted to values $|Z_z|\le 3\textrm{std}(Z_z)$ (yellow line) and shift of average oscillation period in perturbation experiments with delta kicks of strength $\varepsilon=0.5$ in the $z$ direction. (d) Range of values of the $Z_z$ (thin blue lines) and average value (black line) as a function of $\vartheta$. The dashed lines mark three standard deviations.}\label{Fig:Lorenz1}
\end{figure}
\subsection{Lorenz System}\label{Sec:Lorenz}
\begin{figure}[!t]
\setlength{\unitlength}{1cm}
\begin{picture}(4.2,4.2)
\put(0,0){\includegraphics[height=4.4cm]{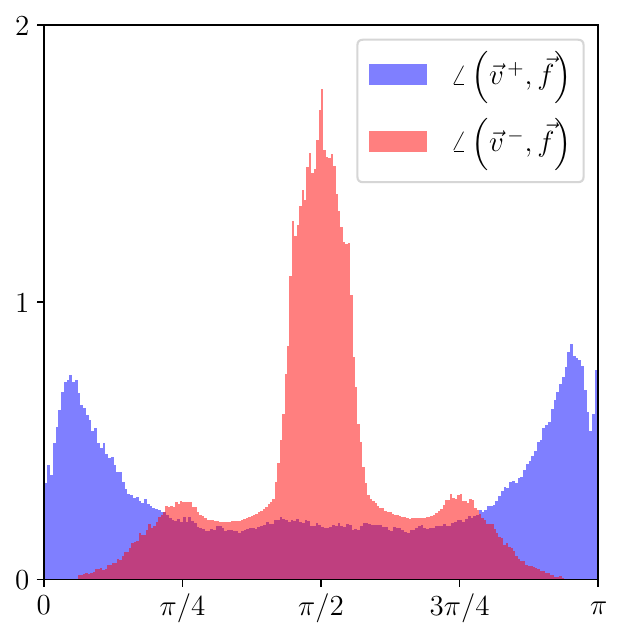}}
\put(-0.2,3.8){\bf (a)}
\end{picture}
\begin{picture}(4.2,4.2)
\put(-0.1,0){\includegraphics[height=4.2cm]{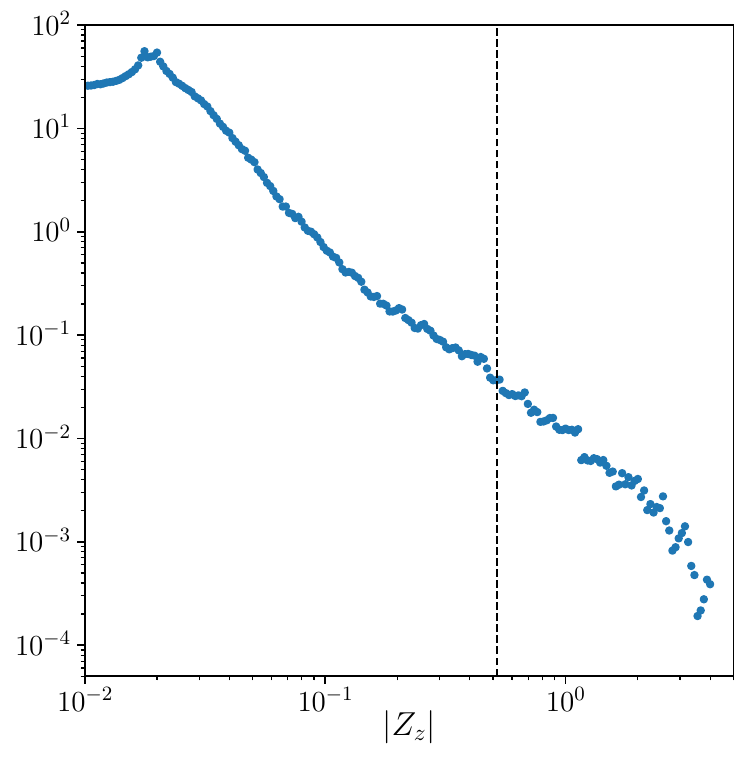}}
\put(-0.1,3.8){\bf (b)}
\end{picture}
\caption{Tangencies between the unstable direction $\vec{v}^{\,+}$ and the flow $\vec{f}$ lead to divergence of the phase sensitivity, which is orthogonal to the stable and unstable directions but is normalized as $\vec{Z}\cdot\vec{f}=1$. (a) Histogram of the angles between the unstable subspace $\vec{v}^{\,+}$ (blue), the stable subspace $\vec{v}^{\,-}$ (red) and the flow $\vec{f}$. (b) Double logarithmic histogram with logarithmic binning of the values $|Z_z|$ shows power law scaling over two orders of magnitude. The dashed line marks three standard deviations.}\label{Fig:Lorenz2}
\end{figure}

Finally we present phase and frequency response in the non-hyperbolic chaotic Lorenz system
\begin{eqnarray}\label{Eq:LorenzSystem}
    \dot x &=& \sigma(y-x) \\
    \dot y &=& x(\rho - z) - y \\
    \dot z &=& xy - \beta z
\end{eqnarray}
We use $\sigma=16$, $\beta=2.0$ and $\rho=28$ where the system is chaotic with Lyapunov exponents $\Lambda^+=0.8$ and $\Lambda^-=-20$. The proto-phase is defined by $R \cos\vartheta_0 = z - z_0$ and $R\sin\vartheta_0 = \sqrt{x^2+y^2}-u_0$ with respect to the fixed point coordinates $u_0 = \sqrt{2\beta(\rho-1)}$ and $z_0 = \rho-1$. We expand the optimized phase around that proto-phase as
\begin{equation}
    \vartheta_{\sigma} = \vartheta_0 + \sum_{k=0}^{5}\sum_{l=0}^{3} \sigma^\pm_{kl}q^\pm_{kl}
\end{equation}
with 
\begin{equation}
    q^+_{kl} = \cos(k\vartheta_0)R^l, \qquad
    q^-_{kl} = \sin(k\vartheta_0)R^l,
\end{equation}
$\sigma^+_{00}=0$ and $\sigma^-_{0l}=0$ and find the coefficients $\sigma^\pm_{ml}$ which minimize the variance of the phase velocity. In Fig.\ref{Fig:Lorenz1}a we show a projection of the Lorenz attractor to the $(x,z)$ coordinates. The points are colored according to 100 intervals of the optimized geometric phase. Red and blue shades, respectively, signify positive and negative average frequency response to perturbations in the $z$ direction, predicted by the $z$-component of the phase sensitivity $\vec{Z}$. In Fig.\ref{Fig:Lorenz1}b we project the chaotic oscillations to the coordinates $\left(\sqrt{x^2+y^2},z\right)$ used in the definition of the proto-phase. In numerical experiments we have performed 1000 kicked perturbations $\varepsilon\vec{p}=0.5\vec{e}_z\delta(t-t_{kick})$ at constant optimized geometric phase after each oscillation and measure the resulting shift of the oscillation period. In Fig.\ref{Fig:Lorenz1}c we compare the frequency response in the perturbation experiments to the average phase sensitivity at that geometric phase predicted by the $z$-component of $\vec{Z}$. Shown are the average phase sensitivity after convolution of $Z_z$ with a narrow Gaussian (black curve) and the average phase sensitivity restricted to values within three standard deviations (yellow curve). Apparently large deviations in $Z_z$ have a strong influence on the predicted average response. The frequency response measured in the perturbation experiments (blue crosses) follow in parts the features of both averages but can also deviate significantly from the predictions. The standard deviation $\textrm{std}(Z_z)=0.17$ is ten times larger than the actual response, and the extreme values seem to follow a power-law over two orders of magnitude (Figs.\ref{Fig:Lorenz1}d and \ref{Fig:Lorenz2}b). The reason for this are frequent near tangencies of the unstable Lyapunov direction and the flow, which can be seen in Fig.\ref{Fig:Lorenz2}a from the distribution of angles (blue histogram). The Lorenz system is an example of a non-hyperbolic chaotic oscillator where our method performs poorly.
\section{Conclusions}
Measuring the frequency response to pulsed perturbations at a given Poincare section is a simple and experimentally viable way to define and measure phase response functions of chaotic oscillators. In this work we have presented a theoretical approach to predict these frequency shifts with the help of co-variant Lyapunov vectors. A phase sensitivity $\vec{Z}=\vec{Z}(\vec{x}_0)$ can be constructed for the points on the chaotic attractor. Time shifts along a chaotic trajectory in response to arbitrary perturbations can be calculated to the linear order of the perturbation strength in the same way as for limit cycle oscillators with Winfree type phase equations Eq.\,\eqref{Eq:LCPhaseResponse}. These time shifts are only exact for a certain perturbed trajectory shadowing the unperturbed trajectory. However, averaging the time shifts for time independent perturbations over the whole attractor can approximate the frequency shift for arbitrary perturbed trajectories. Given the phase sensitivity $\vec{Z}(\vec{x}_0)$ a differentiable geometric phase $\vartheta(\vec{x}_0)$ can be constructed with a gradient $\vec{\nabla}\vartheta$ which approximates the phase sensitivity and minimizes the variance of the phase velocity on the attractor and in its vicinity. We demonstrate our theory with a chaotic electro-chemical oscillator and the chaotic Roessler oscillator, both examples of non-hyperbolic, i.e. non structurally stable systems, where the numerically determined Lyapunov vectors can give good approximations of the linear frequency response. Because of large deviations in the Lyapunov co-vector field $\vec{Z}(\vec{x}_0)$ Frequency response in the non-hyperbolic chaotic Lorenz system is not well predicted. We have also included an example of hyperbolic autonomous oscillations, where the Lyapunov vectors and the phase sensitivity $\vec{Z}(\vec{x})$ can be determined numerically robustly.
\begin{acknowledgments} 
We thank Z. Arai, H.\,Nakao, A.\,Pikovsky and K.\,Takeuchi for valuable discussions. H.K. acknowledges the financial support from MEXT KAKENHI Grant No. 15H05876 and JSPS KAKENHI Grant No. 18K11464.
\end{acknowledgments}
\section*{Data Availability Statement}
The data that support the findings of this study are available from the corresponding author upon reasonable request.
%
%
\bibliography{FreqPRCsBib}
\appendix

\section{}\label{Sec:AppendixA}

We use the method developed by Ginelli et al. \cite{ginelli2007characterizing} to determine the co-variant Lyapunov vectors $\vec{v}^{(k)}(\varphi)$ along a chaotic trajectory $\vec{x}_0(\varphi)$ evolving according to Eq.\,\eqref{Eq:x0_Dyn} on the system attractor. Both, the time step $d\varphi$ forward map $\vec{M}(\vec{x}_0,d\varphi)=\vec{x}_0(\varphi+d\varphi)$ 
\begin{equation}
    \frac{d}{d\varphi}\vec{M} = \vec{f}\left(\vec{M}\right),\qquad \vec{M}(0) = \vec{x}_0(\varphi)
\end{equation}
and its Jacobian matrix $\textrm{J}_M$ with
\begin{equation}
    \frac{d}{d\varphi}\textrm{J}_M = \textrm{J}_f\left(\vec{M}\right)\cdot\textrm{J}_M, \qquad 
    \textrm{J}_M(0) = \mathbf{1}
\end{equation}
are integrated simultaneously by standard RK4 fourth order Runge-Kutta method.
The Lyapunov vectors (except $\vec{v}^{(0)}=\vec{f}$) are normalized $|\vec{v}^{(k)}|=1$ so that $d\vec{v}^{(k)}/d\varphi$ and $\vec{v}^{(k)}$ are orthogonal. With that and from Eq.\,\eqref{Eq:LyapunovEvolution} follow the local Lyapunov exponents
\begin{equation}
    \lambda^{(k)} = \vec{v}^{(k)}\cdot\textrm{J}\vec{v}^{(k)}, \qquad \textrm{(for }|\vec{v}^{(k)}|=1\textrm{).}
\end{equation}
Convergence of the Lyapunov vectors means independence from initial conditions in both forward and backward integration. Choosing two different random initial matrices of Lyapunov vectors, convergence to the co-variant Lyapunov vectors can be monitored.
Given the matrix $\textrm{V}=(\vec{f},\vec{v}^{(1)},\vec{v}^{(2)},\dots)$ of co-variant Lyapunov vectors, the matrix $\textrm{U}=(\vec{Z},\vec{u}^{(1)},\vec{u}^{(2)},\dots)$ of co-variant Lyapunov co-vectors is simply the inverse Matrix of $\textrm{V}$, i.e.
\begin{equation}
    \textrm{U}^\top\textrm{V}=\textrm{V}^{-1}\textrm{V}=\mathbf{1}.
\end{equation}
The Lyapunov co-vectors do not have unit length. Because of biorthonormality, alignment of the Lyapunov vectors, brings $\textrm{V}$ closer to degeneracy and results in large Lyapunov co-vectors. Where Lyapunov vectors, and thus stable, neutrally stable and unstable subspaces become tangential, $\textrm{U}$, and $\vec{Z}$ in particular, is divergent.
\end{document}